\documentclass[11pt,letterpaper,authoryear]{elsarticle}

\usepackage[margin=0.65in]{geometry}
\usepackage{amsmath,amsfonts}
\usepackage{booktabs}
\usepackage{graphicx}
\usepackage{indentfirst}
\usepackage{lmodern}
\usepackage{url}
\usepackage{float}
\usepackage{todonotes}

\newcommand{\articlegraphic}[2][]{%
    \IfFileExists{#2}{%
        \includegraphics[#1]{#2}%
    }{%
        \PackageWarningNoLine{article-source}{Missing figure: #2}%
        \fbox{\parbox[c][0.15\textheight][c]{0.85\linewidth}{%
            \centering\footnotesize Missing figure:\\[0.5em]
            \texttt{\detokenize{#2}}}}%
    }%
}

\begin{document}

\begin{frontmatter}

\title{\textbf{Anchored Identification of Market Regimes for Dynamic Asset Allocation}}

\author[eth]{Antoine Allain}
\ead{aallain@ethz.ch}

\author[kaust]{Florian Kogelbauer}
\ead{florian.kogelbauer@kaust.edu.sa}

\affiliation[eth]{
    organization={ETH Z\"urich, Department of Mathematics},
    addressline={R\"amistrasse 101},
    city={Z\"urich},
    postcode={8092},
    country={Switzerland}
}

\affiliation[kaust]{
    organization={King Abdullah University of Science and Technology (KAUST)},
    addressline={CEMSE Division},
    city={Thuwal},
    postcode={23955-6900},
    country={Saudi Arabia}
}

\begin{abstract}
\noindent
 A portfolio manager who allocates assets depending on market regimes needs the regimes to retain their economic meaning when the underlying model is re-estimated. We develop a semi-supervised procedure that uses economic templates and fourteen historical months selected as clear examples of crisis, inflation, expansion, or fragile market conditions to guide the identification of these regimes. Using twelve monthly macro-financial indicators from 1971 to 2026, we evaluate the procedure in Gaussian and Student-$t$ mixture models and hidden Markov models, with Hungarian matching and constrained expectation-maximization as benchmarks. From January 1990 through April 2026, the same three-asset allocation rule applied to the Student-$t$ hidden Markov model produced annualized sample Sharpe ratios of 1.19 under anchored identification, 0.98 under Hungarian matching, and 0.83 under constrained EM. The anchored strategy compounded at 9.16\% per year with a maximum drawdown of 17.08\%, compared with 7.01\% growth and a 33.15\% drawdown for equal weighting of the same three assets. Buy-and-hold US equities earned more, at 11.07\% per year, and experienced a maximum drawdown of 50.95\%.
\end{abstract}


\end{frontmatter}

\pagebreak

\section{Introduction}

A portfolio manager who aims at reducing equity exposure still has to decide where to reinvest. Treasuries may be attractive when credit spreads widen and investors seek safety, while commodities may offer some protection during inflation. A regime model can help to distinguish between these situations, provided that a state associated with a certain macroeconomic state continues to carry that meaning as the model is refitted. We investigate whether a small set of clearly identifiable historical episodes can guide the estimation of macro-financial regimes and whether the resulting distinctions improve portfolio allocation.\\
The aforementioned difficulty arises because adding observations to the estimation sample can change what a fitted state represents. An unsupervised model may recover the same states but in a different order \citep{stephens2000label}, i.e., a permutation that leaves portfolio holdings unchanged whenever each state is matched to its original portfolio. The harder problem emerges as refitting can change the clusters themselves: a state initially characterized by widening credit spreads, for example, may later absorb months driven mainly by rising commodities. Retaining the Crisis label would then direct the portfolio toward Treasuries under conditions for which the manager might have preferred commodities.\\
To overcome these difficulties, we propose a semi-supervised procedure that is based on fourteen retrospectively selected months as historical reference points for our four regimes: Crisis, Inflation, Expansion, and Fragile. These dated examples complement economic templates that describe the cross-asset behavior associated with each regime. Building on semi-supervised clustering and anchored mixture models \citep{basu2002semi,kunkel2020anchored}, we incorporate these dates and templates into an expectation-maximization (EM) procedure \citep{dempster1977maximum}. We call the procedure ordered-state, anchored, common-shape identification EM (OACI-EM). We compare this method against two benchmarks. The first benchmark fits the model without the economic templates, then assigns the fitted states to the templates by Hungarian matching \citep{kuhn1955hungarian,munkres1957algorithms}. The second benchmark uses constrained EM. Parameter restrictions have been used in EM estimation of Gaussian mixtures \citep{hathaway1986constrained}. Here the bounds encode the economic characteristics assigned to each regime.\\
Regime-switching models are standard tools for studying economic activity \citep{hamilton1989new} and asset returns \citep{ang2012regime}. In portfolio management, they have been used to time equities \citep{bulla2011markov}, allocate jointly across stocks and bonds \citep{guidolin2007asset}, optimize portfolios over hidden states \citep{baeKimMulvey2014Varied}, and tilt weights based on views about regime probabilities \citep{elkamhi2023portfolio}. Other work also uses trend filtering to flag downside risk in endowment portfolios \citep{mulvey2016identifying}. Regimes are usually named to fit the specific allocation task: for equity timing, \citet{shu2024statistical} label the state with the higher in-sample cumulative excess return as bullish, while \citet{shu2025dynamic} fit separate bull and bear states for each asset and \citet{yu2026allocation} define regimes by the relative performance of investment strategies. By contrast, allocating across equities, Treasuries, and commodities requires a single shared set of market regimes rather than separate asset-specific labels.\\
We study twelve monthly macro-financial indicators from March 1971 through April 2026 using Gaussian and Student-$t$ mixtures and hidden Markov models \citep{bulla2011hidden}. At each end of the month, we assess current conditions with a model estimated on earlier observations, allowing a historical anchor date to enter estimation only after a twelve-month lag. The resulting regime scores determine the following month's holdings. In backtests from January 1990 through April 2026, we first hold the allocation rule fixed across all models so that differences in holdings only come from the estimated regimes, before examining how portfolio construction itself affects the results. For the seven-asset comparison, each regime has a portfolio estimated before the backtest, and the scores determine how much to invest in each portfolio. A further experiment holds the regime signal fixed and updates the portfolios monthly, using historical scores paired with returns from either the same month or the following month.\\
Overall, semi-supervised estimation produces regimes that are easier to interpret macroeconomically and more effective for asset allocation. In the Student-$t$ models, hard parameter bounds collapse most of the timeline into Expansion and leave Fragile almost unused, while Hungarian matching blurs the economic roles of the states, with a higher recession frequency in Fragile than in Crisis. Guiding estimation with historical anchors and economic templates keeps all four regimes active and consistent with external macroeconomic data even after excluding the anchor months, with recession frequency highest in Crisis and average inflation highest in Inflation. In the out-of-sample backtests, for the Student-$t$ hidden Markov model under the three-asset rule (equities in Expansion, commodities in Inflation, and Treasuries in Crisis and Fragile), semi-supervised identification raises the sample Sharpe ratio from 0.83 under constrained estimation to 1.19. The anchored strategy also compounds faster than equal weighting of the same three assets (9.16\% versus 7.01\% per year), while nearly halving the maximum drawdown (17.08\% versus 33.15\%). Under the defensive bond-floor rule, the anchored Student-$t$ hidden Markov model also achieves the highest sample Sharpe ratio among the twelve model specifications.

\section{Data}

Our panel comprises twelve monthly macro-financial indicators from March 1971 through April 2026. We choose indicators that help distinguish episodes with similar equity returns but different conditions in other markets. For example, widening credit spreads and rising implied volatility can signal financial stress even when equity returns remain calm. Treasury and commodity returns show how two potential alternatives to equities behave during those episodes, while the term spread and the inflation-linked series provide additional information on interest rates and inflation. Table~\ref{tab:factor-panel} lists the series and their transformations.\\
Before March 1997, the inflation-linked indicator uses the latest released change in year-on-year US consumer price index (CPI) inflation, in percentage points, calculated from revised CPI observations, before switching to monthly Treasury Inflation-Protected Securities (TIPS) index returns. Although this splice changes both economic meaning and units, the series is used only to estimate the regimes (both pre-1990 and in the expanding walk-forward fits) and is never held in any portfolio. Appendix~\ref{app:data-provenance} details the release rule and the remaining historical substitutes.\\
A fixed regime portfolio is a set of asset weights estimated once for a given regime and kept constant throughout the backtest. The commodity returns used to estimate these portfolios include Bloomberg's historical reconstruction. For the 1960-1990 extension, Bloomberg applied its 1998 target weights to the contracts eligible in each historical year. MSCI World price returns stand in for emerging markets through 1987, and a trend strategy using commodities and Treasuries stands in for the trend index before 2000. For portfolio estimation only, we add an assumed 4\% annual dividend yield to early US equity price returns.\\
To place the twelve series on a comparable scale, we standardize each monthly change using its exponentially weighted mean and bias-adjusted standard deviation computed through the previous month, with a 60-month span and a minimum history of 36 observations. For numerical stability and outlier control, we floor the standard deviation at $10^{-8}$ and clip the resulting $z$-scores at $\pm4$ (see Appendix~\ref{app:data-provenance} for the exact recursions). Because each indicator is centered on its trailing average, a positive standardized value simply means the monthly change sits above recent norms, even when the raw change is negative.

\begin{table}[ht]
\centering
\setlength{\tabcolsep}{5pt}
\resizebox{\textwidth}{!}{%
\begin{tabular}{@{}llll@{}}
\toprule
Indicator & Ticker & Asset class & Transformation \\
\midrule
S\&P 500 total return
& \texttt{SPXT} & Equities & Return \\
US Treasury total return
& \texttt{LUATTRUU} & Rates & Return \\
10Y-3M term spread
& \texttt{YC\_10Y\_3M} & Rates & First difference \\
US high-yield corporate bond return
& \texttt{LF98TRUU} & Credit & Return \\
US aggregate credit option-adjusted spread
& \texttt{LUACOAS} & Credit & First difference \\
Cboe Volatility Index (VIX)
& \texttt{VIX} & Volatility & Percent change \\
Commodity total return
& \texttt{BCOMTR} & Commodities & Return \\
US TIPS total return
& \texttt{BCIT1T} & Inflation-linked & Release-dated proxy, then return \\
US dollar index
& \texttt{DXY} & FX & Return \\
MSCI Emerging Markets
& \texttt{MXEF} & Emerging markets & Return \\
SG trend-following index
& \texttt{NEIXCTAT} & Trend-following & Return \\
Copper/gold relative return
& \texttt{CU\_AU\_RATIO} & Relative value & Log-return differential \\
\bottomrule
\end{tabular}%
}
\caption{Monthly macro-financial indicators. Transformations refer to the native series. Historical substitutes, their start dates, and return conventions are documented in Appendix~\ref{app:data-provenance}.}
\label{tab:factor-panel}
\end{table}

\section{Methodology}

Let $x_t\in\mathbb{R}^d$ denote the vector of standardized indicators for month $t$, with $d=12$, and let $z_t\in\{0,\ldots,K-1\}$ denote the latent state, with $K=4$. We fit Gaussian and Student-$t$ mixture models (GMM and SMM) and hidden Markov models with Gaussian and Student-$t$ emissions (GHMM and SHMM) as follows. Given parameters estimated on observations preceding month $t$, a mixture model assigns the state probabilities from $x_t$ alone. In a hidden Markov model, the states follow a first-order Markov chain, so the filtered probability of each state also reflects the evidence from preceding months. All four models use a diagonal scale matrix for $x_t \mid z_t$: in the Gaussian models this is simply the covariance, and in the Student-$t$ models, where we fix the degrees of freedom at $\nu=6$, the covariance is $\nu/(\nu-2)=1.5$ times larger. Because the Student-$t$ distribution has heavier tails, its estimation updates also downweight observations far from the fitted state mean. For the full equations and estimation routines, we refer to Appendix~\ref{app:baseline-models}.\\
Permuting the state indices of an unconstrained model leaves its likelihood unchanged \citep{stephens2000label} and has no effect on portfolio holdings as long as each state stays paired with the same portfolio. Economic identification, however, matters because refitting can change which months fall into each state, so a portfolio chosen for that state earlier may end up trading under different market conditions. We therefore define Crisis, Inflation, Expansion, and Fragile through explicit indicator patterns and assign them the fixed indices $0$, $1$, $2$, and $3$.\\
Both \textbf{Crisis} and \textbf{Fragile} carry positive signs on changes in the VIX and credit spreads, and differ in how equities behave: Crisis combines this stress with below-average US equity and high-yield bond returns, while Fragile combines it with above-average US and emerging-market equity returns. \textbf{Expansion} reverses the Crisis pattern, with above-average US equity and high-yield returns alongside below-average changes in the VIX and credit spreads. \textbf{Inflation} puts positive signs on commodity returns, the inflation-linked indicator, and changes in the term spread. Because all signs refer to standardized indicators, an above-average reading does not necessarily mean a positive raw return.\\
For each regime $r$ and indicator $j$, the template consists of a pair of signs $s_{r,j}^{\mu}, s_{r,j}^{\sigma} \in \{-1,0,+1\}$ specifying the intended directions of the state mean and dispersion. In the template score, $+1$ favors a larger fitted value, $-1$ favors a smaller one, and $0$ omits that term. Our semi-supervised procedure combines these templates with fourteen retrospectively selected historical months whose regime labels enter estimation as anchors. We compare it with Hungarian matching, which assigns names after estimation, and constrained expectation-maximization, which imposes regime-specific parameter bounds during estimation. The full templates and anchor dates are listed in Tables~\ref{tab:sign-templates} and~\ref{tab:anchor-calendar}, and Section~\ref{sec:external-validation} checks how closely the inferred regimes match macroeconomic data.

\subsection{The Semi-supervised identification (OACI-EM)}

The economic templates describe the behavior we associate with each regime, while fourteen historical months provide examples for which we specify the regime directly. OACI-EM uses these examples as the labeled observations in a semi-supervised fit. These dates were chosen retrospectively, and each anchor enters training only twelve months after its date. We use October and November 2008, when financial markets remained under acute stress after the collapse of Lehman Brothers, and March 2020, when markets fell sharply at the onset of the COVID-19 pandemic, as hard Crisis anchors. Once eligible, each of these months is assigned to Crisis with probability one during estimation. The remaining eleven dates serve as soft anchors. They include September 2008 for Crisis, September 1974 and March, June, and September 2022 for Inflation, June 1997, June 2005, and June 2016 for Expansion, and August 2011, August 2015, and December 2018 for Fragile.\\
Estimation starts by matching the $K$-means centroids one-to-one to the mean templates, fixing which component represents each regime before the first update. At each iteration of EM, the expectation step (E-step) calculates the probability that each training observation belongs to each state under the current parameter estimates. For a mixture, write $\gamma_{s,k}$ for this probability in month $s$ and $\gamma_s=(\gamma_{s,0},\ldots,\gamma_{s,K-1})$ for the vector of responsibilities. If month $s$ is an eligible anchor with designated regime $k^\star$, we adjust this vector before updating the parameters. At a soft anchor, we use posterior regularization \citep{ganchev2010posterior} to find the smallest adjustment in Kullback-Leibler divergence that gives $k^\star$ a responsibility of at least $b=0.90$. If we let $\Delta^{K-1}=\{p\in[0,1]^K:\sum_{k=0}^{K-1}p_k=1\}$ denote the set of probability vectors over the $K$ regimes, the adjusted responsibilities solve
\begin{equation}
\gamma_s^{\mathrm{PR}}
=
\operatorname*{arg\,min}_{p\in\Delta^{K-1}}
\mathrm{KL}(p\|\gamma_s)
\quad\text{subject to}\quad p_{k^\star}\ge b.
\end{equation}
This projection has a closed-form solution: $\gamma_s$ stays unchanged if $\gamma_{s,k^\star}\ge b$, and otherwise the target entry is raised to $b$ while the remaining entries are rescaled proportionally to sum to $1-b$. Hard Crisis anchors correspond to the boundary case $b=1$, placing full probability mass on Crisis.\\
In a hidden Markov model, the state at an anchor month is coupled to the transitions entering and leaving that date, so we apply the labels to the log-emission potentials before running the forward-backward algorithm. Hard anchors allow only the designated state, while soft anchors iteratively raise the target state's log-emission potential toward a $0.90$ marginal probability. Because this update stops once the target is met within numerical tolerance or the iteration limit is reached, a soft anchor can finish below $0.90$. A final forward-backward pass then gives the state and transition probabilities under the same adjusted distribution.\\
The parameter update combines the evidence from these training probabilities with the economic templates. For each mean coordinate subject to shrinkage, we average the estimate from the observations with its template target, giving the data more weight as the total observation weight for the state increases. The default template weight is $100$ and rises to $400$ for changes in the VIX and credit spreads in the Fragile regime. We estimate the diagonal scale entries from residuals about these template-adjusted means before imposing six bounds on the means themselves. In standardized units, with $\varepsilon=0.25$, the Crisis mean for US equity returns must be at most $-2\varepsilon$ and its mean change in the VIX at least $\varepsilon$. For Inflation, the mean changes in the VIX and credit spreads must be nonpositive, while the means for commodity returns and the inflation-linked indicator must each be at least $2.5\varepsilon$.\\
We give the diagonal scale matrices a common shape across regimes, so that each state has its own overall scale while the ratios between indicators' scale entries are shared \citep{celeux1995gaussian}. Let $\sigma_k=(\sigma_{k,1},\ldots,\sigma_{k,d})^\top$, where
$\sigma_{k,j}$ is the square root of the $j$th diagonal entry of the
scale matrix in state $k$. Writing $\sigma_k^2$ for the vector of
componentwise squares, we have, before numerical flooring,
\begin{equation}
\mathrm{diag}(\sigma_k^2)=\tau_k\,\mathrm{diag}(v),
\qquad k=0,\ldots,K-1,
\end{equation}
where $v_j$ determines the relative dispersion of indicator $j$ and $\tau_k$ determines the overall scale of state $k$. We estimate both from weighted residual moments and impose $\tau_{\mathrm{Crisis}}\geq\tau_k$ for every $k$, pooling state scales when the unrestricted estimates violate this ordering. For mixtures, the mixing-weight update also redistributes excess mass from Fragile and then from Inflation. Appendix~\ref{app:oaci-em-details} gives the coordinate weights, redistribution rule, and numerical details.\\
For a new month $t$, which receives no calendar label, the fitted model produces a mixture responsibility or filtered HMM probability $p(k)$ for each regime $k$. We use its observed indicators to measure agreement with the mean template over the coordinates $J_k^\mu=\{j:s_{k,j}^{\mu}\ne0\}$ on which that template specifies a direction. The mean signed alignment is
\begin{equation*}
\mathrm{align}(x_t,s_k^\mu)
=
\frac{1}{|J_k^\mu|}
\sum_{j\in J_k^\mu}s_{k,j}^{\mu}x_{t,j}.
\end{equation*}
A larger value gives the regime greater weight in the adjusted score
\begin{equation}
\label{eq:oaci-tilt}
p^{\mathrm{adj}}(k)
=
\frac{p(k)\exp\{\beta\,\mathrm{align}(x_t,s_k^\mu)\}}
     {\sum_{r=0}^{K-1}p(r)\exp\{\beta\,\mathrm{align}(x_t,s_r^\mu)\}},
\qquad \beta=0.4.
\end{equation}
The historical labels therefore enter the score through the fitted parameters, while the final adjustment depends on the indicators observed in month $t$.

\subsection{Benchmark identification rules}

We compare our semi-supervised regime identification method to two other approaches. Neither benchmark uses historical anchor dates: Hungarian matching fits the model freely and names the states afterward from the templates, and constrained EM fixes each state's label at initialization by bounding its parameters during estimation.

\paragraph{Hungarian matching.}
We fit the underlying mixture or HMM as described in Appendix~\ref{app:baseline-models} and compare each fitted component $i$ with each economic template $r$. For indicator $j$, let $\mu_{i,j}$ denote its conditional mean in component $i$ and $\sigma_{i,j}$ its conditional standard deviation for Gaussian emissions or the square root of its diagonal scale entry for Student-$t$ emissions. The agreement score is
\begin{equation}
\label{eq:hungarian-score}
S_{i,r}
=
\sum_{j=1}^{d}w_j b_{r,j}
\bigl(s_{r,j}^{\mu}\mu_{i,j}
     +s_{r,j}^{\sigma}\sigma_{i,j}\bigr),
\qquad i,r\in\{0,\ldots,K-1\}.
\end{equation}
The inverse-scale weight $w_j$ accounts for differences in dispersion across indicators, while the multiplier $b_{r,j}\geq1$ up-weights key indicators for regime $r$. Over the set $\mathcal{S}_K$ of permutations of the $K$ labels, the Hungarian algorithm \citep{kuhn1955hungarian,munkres1957algorithms} finds the assignment with the highest total score,
\begin{equation}
\rho^\star
\in
\operatorname*{arg\,max}_{\rho\in\mathcal{S}_K}
\sum_{i=0}^{K-1}S_{i,\rho(i)}.
\end{equation}
Component $i$ receives the name $\rho^\star(i)$ and carries its fitted parameters and probabilities with it. When two assignments have nearly equal total scores, a small change in the fitted components can reverse their ordering at the next refit. Appendix~\ref{app:hungarian-details} gives the scoring weights.

\paragraph{Constrained EM.}
Constrained EM restricts the means and dispersions that each named state can attain. For regime $k$, let $J_k$ denote the coordinates with bounds on the mean and let $[\ell_{k,j},u_{k,j}]$ be the admissible interval for coordinate $j$. At $K$-means initialization and after each subsequent location update, we project the unrestricted estimate $\widehat{\mu}_k$ onto the feasible set $C_k$,
\begin{equation}
\label{eq:constrained-box}
C_k =
\bigl\{m\in\mathbb{R}^d:
\ell_{k,j}\le m_j\le u_{k,j}
\text{ for }j\in J_k\bigr\},
\qquad
\mu_k^{\mathrm{proj}}=\Pi_{C_k}(\widehat{\mu}_k),
\end{equation}
which simply clips each constrained coordinate to its admissible interval. With $\varepsilon=0.25$ in standardized units, the bounds distinguish Crisis, which requires weak equity returns and elevated financial stress, from Fragile, which allows a mean equity return in $[0,\varepsilon]$ alongside elevated changes in the VIX and credit spreads. The Inflation bounds require elevated means for commodity returns and the inflation-linked indicator, while Expansion combines nonnegative mean equity and high-yield returns with nonpositive mean changes in the VIX and credit spreads.\\
After projecting the mean, we estimate the diagonal Gaussian variance or Student-$t$ scale from residuals about that feasible location and clip each entry to a regime-specific interval (listed in full in Appendix~\ref{app:constrained-em-details}). Specifically, Crisis sets dispersion floors on equities, the VIX, and credit spreads, Fragile tightly caps equity dispersion, and Inflation and Expansion face tighter overall ceilings. Mixture weights and HMM transitions retain their standard updates, and both benchmarks score a new observation directly from the fitted model probabilities.

\subsection{Walk-forward protocol}
\label{sec:walk-forward}

Fitting a clustering model once on the full sample introduces look-ahead bias, because future observations help decide which earlier months belong together. To reflect real-time investing, we instead refit each model through the preceding month and classify each new observation as it arrives. We generate 436 out-of-sample month-end classifications from January 1990 through April 2026 (with the March 2026 signal setting the final portfolio return in April). At each test month $t$, every model is fitted using only past observations,
\begin{equation}
\mathcal{D}_t=\{x_s:s<t\}.
\end{equation}
Mixtures and HMMs are refitted monthly on expanding samples (requiring at least 40 prior raw months before standardization).\\
The fourteen anchors were selected retrospectively and are used in estimation only once they are in the training window, with a twelve-month lag before each anchor becomes eligible. Under this rule, September 1974 (Inflation) is the only anchor available for the pre-1990 fit, while the first anchors for Expansion, Crisis, and Fragile become eligible in July 1998, October 2009, and September 2012, respectively.\\
At month-end $t$, an HMM updates its predictive state distribution with $x_t$ to produce the filtered probabilities $\Pr(z_t=k\mid x_{1:t})$, while a mixture computes posterior responsibilities from $x_t$ alone. Under Hungarian matching and Constrained EM, these probabilities form the regime score vector $g_t$ unchanged, and under OACI-EM, we first apply the template adjustment in \eqref{eq:oaci-tilt}. Every mixture and HMM then blends $g_t$ with its trailing exponential moving average:
\begin{equation}
\label{eq:signal-blend}
\tilde g_t=w\,g_t+(1-w)e_t,
\qquad
e_t=(1-\alpha_S)e_{t-1}+\alpha_S g_t,
\end{equation}
with $w=0.35$, $S=8$, and $\alpha_S=2/(S+1)$, initializing $e_t$ at the first out-of-sample score. These smoothed scores $\tilde g_t$ then serve as the weights on the four regime portfolios.\\
To construct the discrete trading signal $\hat z_t^{\mathrm{sig}}$, we define the candidate regime $\hat z_t^{\mathrm{cand}}=\arg\max_k\tilde g_t(k)$, initialize $\hat z_t^{\mathrm{sig}}$ with the first candidate, and switch only after four completed months in the current regime when the candidate's smoothed score is at least $0.40$. The signal overrides this dwell rule for Crisis, entering immediately once the unsmoothed Crisis score reaches $0.80$ and exiting to the current candidate as soon as it falls below $0.20$. These switching rules affect only $\hat z_t^{\mathrm{sig}}$ and leave the blended scores $\tilde g_t$ unchanged for the main portfolio backtests. Although pre-1990 calibration selected no post-processing for HMMs, we apply the same blend and switching rules to both families so their signals are directly comparable.

\section{Results}

We examine the regimes identified at each month-end from January 1990 through April 2026. Each timeline shows the discrete trading signal $\hat z_t^{\mathrm{sig}}$, the economic label obtained after smoothing the regime scores and applying the switching rules. We measure persistence through spell duration, the number of consecutive months spent in the same regime, which reflects both the fitted model and the switching rules. A change of label can occur because the latest observation points to a different regime or because a refit changes the economic conditions associated with a state.\\
We first examine the macroeconomic conditions associated with the four labels, then use the timelines to compare how the semi-supervised procedure and the two benchmarks classify successive months.

\subsection{Macroeconomic characteristics of the inferred regimes}
\label{sec:external-validation}

Table~\ref{tab:shmm-occupancy} holds the model family fixed and reports the classifications obtained under each identification rule. To assess the economic meaning of the labels beyond the anchor dates, we exclude them from the macroeconomic statistics for all three methods. For each regime, the recession percentage answers the question of how often a month carrying that label falls within a US recession dated by the National Bureau of Economic Research (NBER). We also report average year-on-year growth in the consumer price index (CPI) and industrial production (IP).

\begin{table}[ht]
\centering
\vspace{0.4em}
\small
\begin{tabular}{@{}llrrrrr@{}}
\toprule
Method & Regime & Share (\%) & Duration (months) & NBER (\%) & CPI (\%) & IP (\%) \\
\midrule
OACI-EM & Crisis    & 22.5 & 2.4 & 15.9 & 2.8 (88)  & 0.3 (88) \\
        & Inflation & 12.2 & 4.4 &  7.5 & 3.2 (53)  & 4.4 (45) \\
        & Expansion & 58.0 & 5.1 &  5.6 & 2.5 (249) & 1.3 (246) \\
        & Fragile   &  7.3 & 3.2 &  0.0 & 2.9 (32)  & 2.9 (32) \\
\addlinespace
Hungarian & Crisis    & 14.4 & 1.5 & 15.1 & 3.1 (53)  &  2.2 (52) \\
          & Inflation & 30.3 & 4.7 &  3.8 & 2.5 (131) &  1.4 (124) \\
          & Expansion & 36.5 & 4.5 &  3.2 & 2.6 (156) &  2.6 (157) \\
          & Fragile   & 18.8 & 3.7 & 17.1 & 2.8 (82)  & $-0.8$ (78) \\
\addlinespace
Constrained EM & Crisis    &  4.8 &  1.5 & 25.0 & 2.2 (16)  & $-0.4$ (15) \\
               & Inflation & 15.1 &  4.1 & 12.5 & 3.8 (64)  &  2.6 (56) \\
               & Expansion & 79.1 & 11.9 &  5.9 & 2.5 (338) &  1.5 (336) \\
               & Fragile   &  0.9 &  4.0 &  0.0 & 1.6 (4)   &  2.0 (4) \\
\bottomrule
\end{tabular}
\caption{Macroeconomic profile of the Student-$t$ HMM regimes, January 1990 to April 2026. Excluding the thirteen anchor months in this period, NBER shows the share of each regime's months falling in a recession, while CPI and IP report mean year-on-year growth rates with observation counts in parentheses (data sources in Appendix~\ref{app:data-provenance}).}
\label{tab:shmm-occupancy}
\end{table}

\subsection{Semi-supervised identification with OACI-EM}

\begin{figure}[H]
\centering
\articlegraphic[width=\textwidth]{ 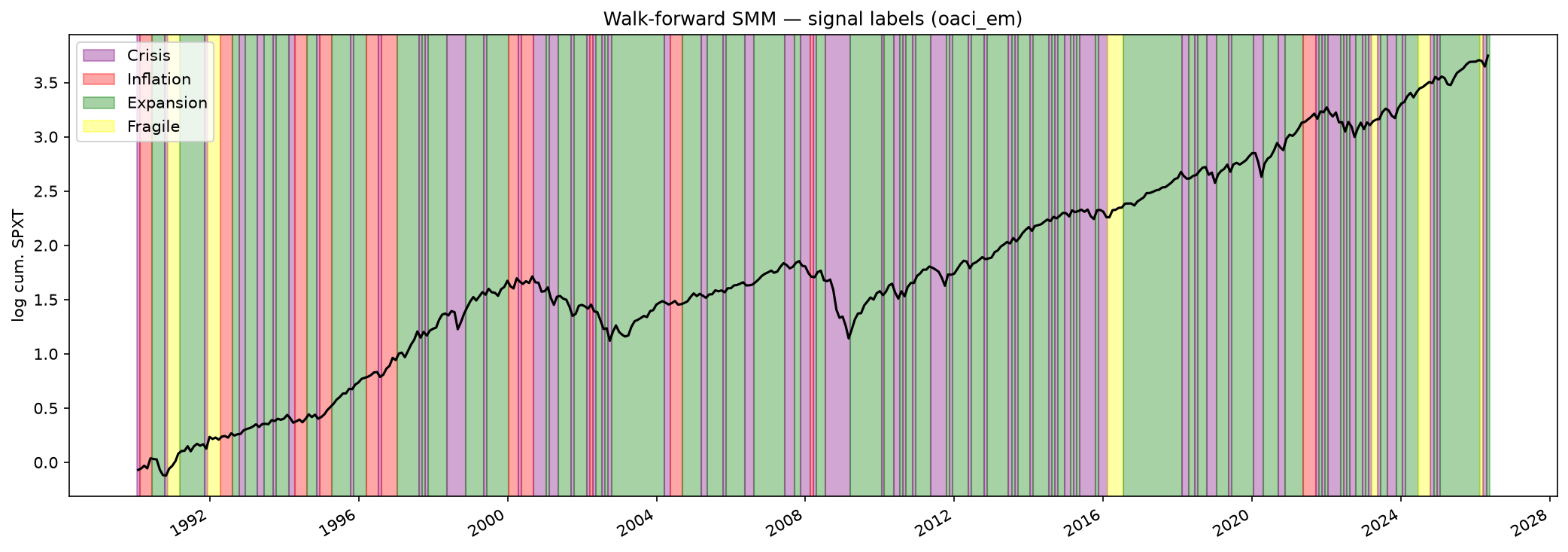}
\caption{Walk-forward classifications for the Student-$t$ mixture under OACI-EM, January 1990 to April 2026. Shading identifies Crisis (purple), Inflation (red), Expansion (green), and Fragile (yellow). The black line shows the log cumulative S\&P~500 total return.}
\label{fig:oaci-timelines}
\end{figure}

\begin{figure}[H]
\centering
\articlegraphic[width=\textwidth]{ 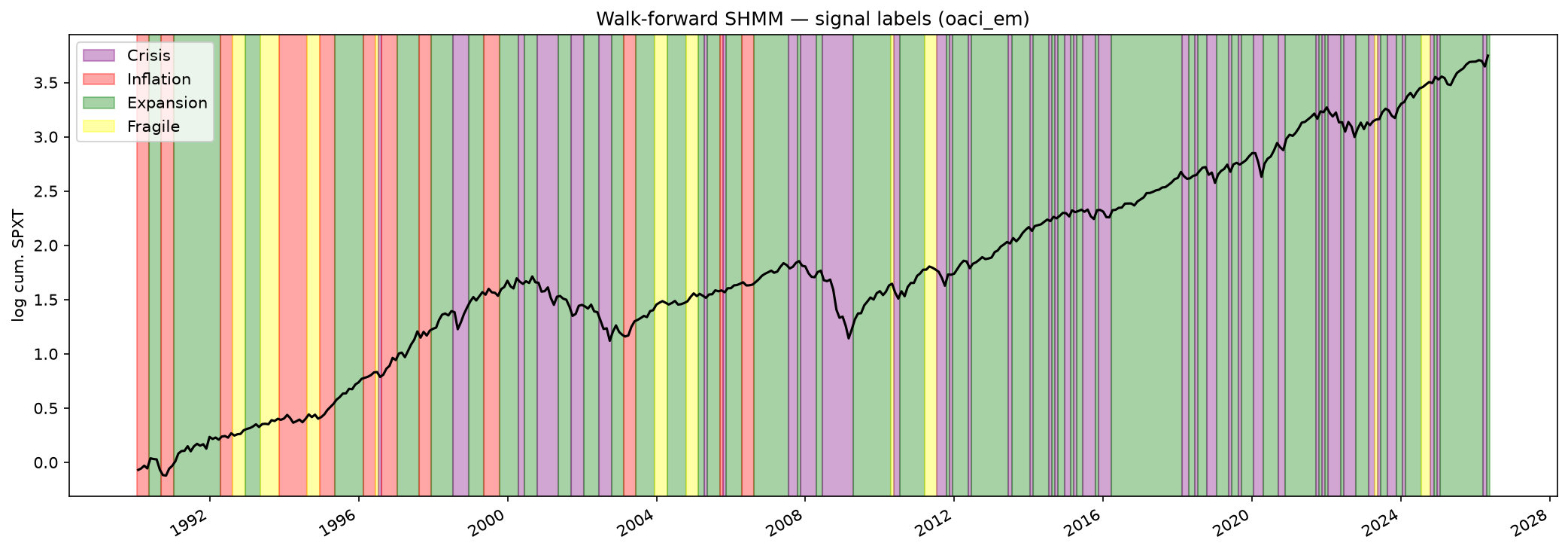}
\caption{Walk-forward classifications for the Student-$t$ HMM under OACI-EM, January 1990 to April 2026.}
\label{fig:shmm-oaci}
\end{figure}

Figures~\ref{fig:oaci-timelines} and~\ref{fig:shmm-oaci} show the anchored Student-$t$ timelines: Expansion covers 58.7\% of months in the SMM and 58.0\% in the SHMM, while Crisis recurs in shorter episodes (27.1\% in the SMM and 22.5\% in the SHMM, where Crisis spells last 2.4 months on average). Both models use the same smoothing and switching rules, yet the SHMM switches labels 112 times across the 435 monthly transitions compared with 145 for the SMM. The anchored Gaussian mixture and HMM likewise keep all four regimes active (Appendix~\ref{app:additional-timelines}).

Because each historical anchor enters training only after a twelve-month lag, its initial out-of-sample classification provides a useful check. On those first passes, the semi-supervised SHMM recovered seven of the thirteen evaluation-period labels, matching all four Crisis and all three Expansion months while assigning the three Fragile and three 2022 Inflation months to Crisis. Smoothing explains the miss in March 2022: the template-adjusted score favored Inflation, and only after smoothing did the top score and discrete signal shift to Crisis. In June and September 2022, Crisis already led before smoothing and remained the selected regime throughout.

\subsection{Benchmark identification rules}

\begin{figure}[H]
\centering
\articlegraphic[width=\textwidth]{ 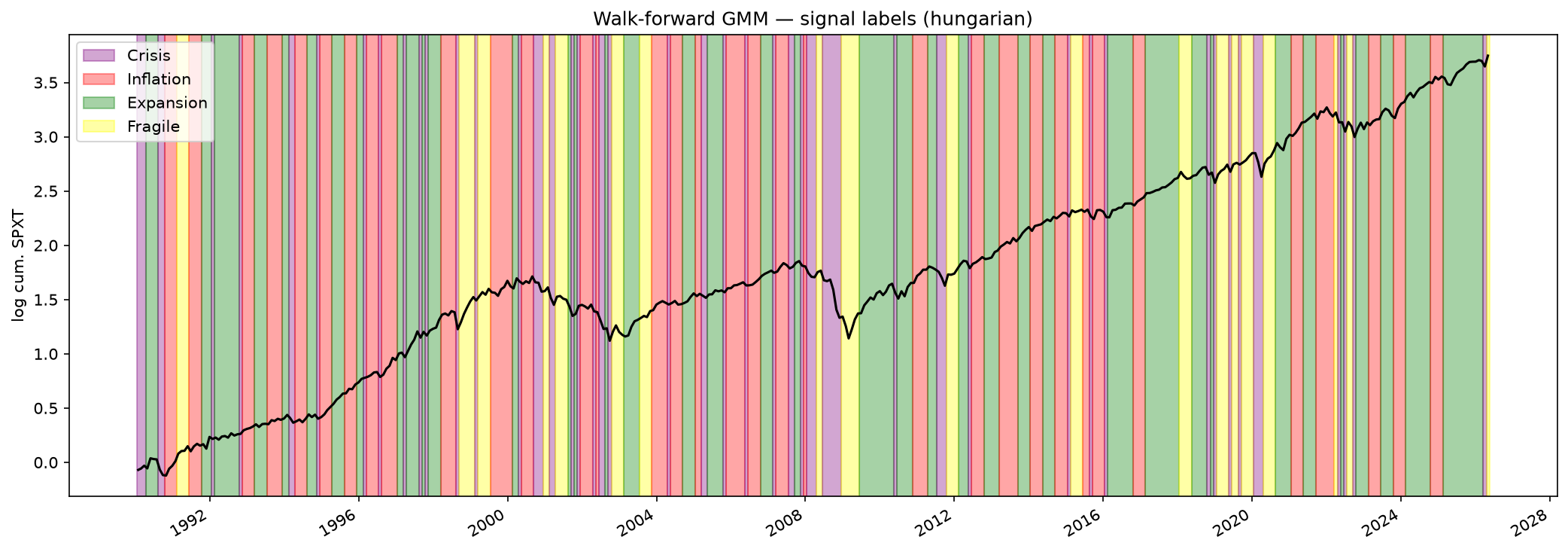}\\[0.6em]
\articlegraphic[width=\textwidth]{ 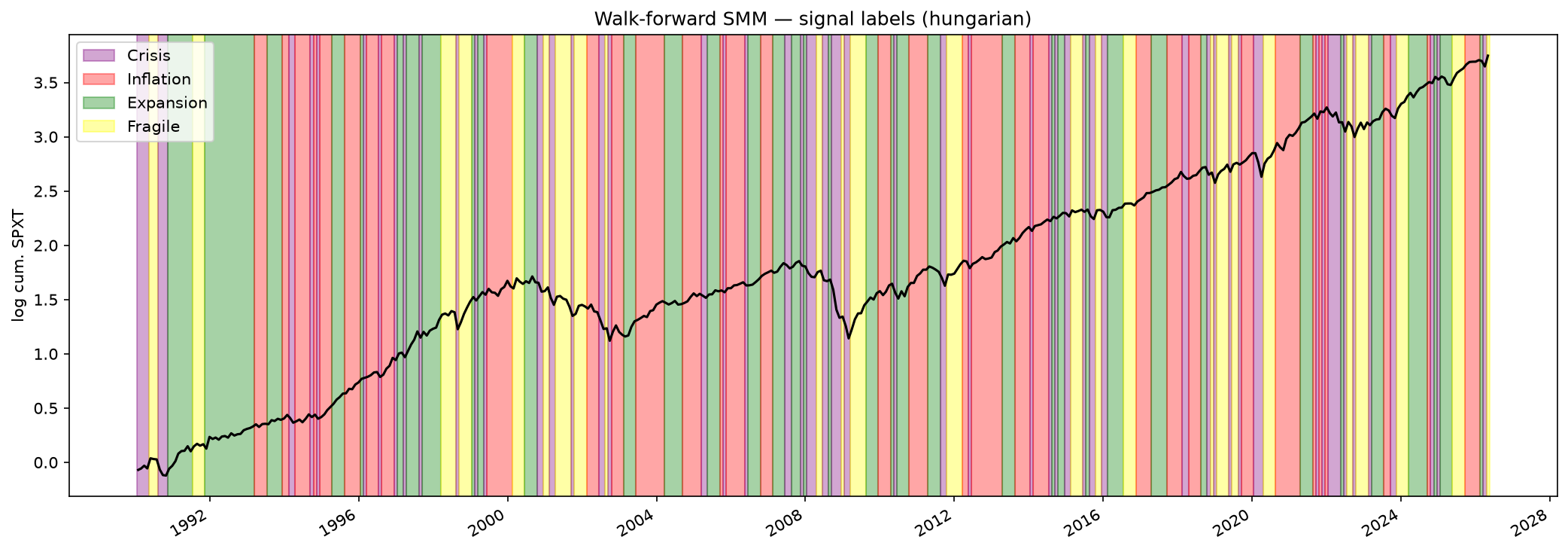}
\caption{Walk-forward classifications under Hungarian matching, January 1990 to April 2026. The upper panel shows the Gaussian mixture and the lower panel the Student-$t$ mixture. Colors and the equity return series are as in Figure~\ref{fig:oaci-timelines}.}
\label{fig:hungarian-mixtures}
\end{figure}

\paragraph{Hungarian matching.}
Under Hungarian matching, both mixtures keep all four labels active (Figure~\ref{fig:hungarian-mixtures}), while the HMMs switch less often (Appendix~\ref{app:additional-timelines}): across the 435 monthly transitions, the GMM and SMM change label 136 and 147 times, compared with 118 and 126 times for the GHMM and SHMM. Within each family, the Gaussian model switches less often because it does not downweight extreme months, leading to a cleaner contrast for the template score than in the heavy-tailed Student-$t$ fit. In the Hungarian SHMM, Expansion and Inflation together cover roughly two thirds of the sample, averaging about four and a half months per spell, while Crisis is the shortest-lived state at 1.5 months on average.

\begin{figure}[H]
\centering
\articlegraphic[width=\textwidth]{ 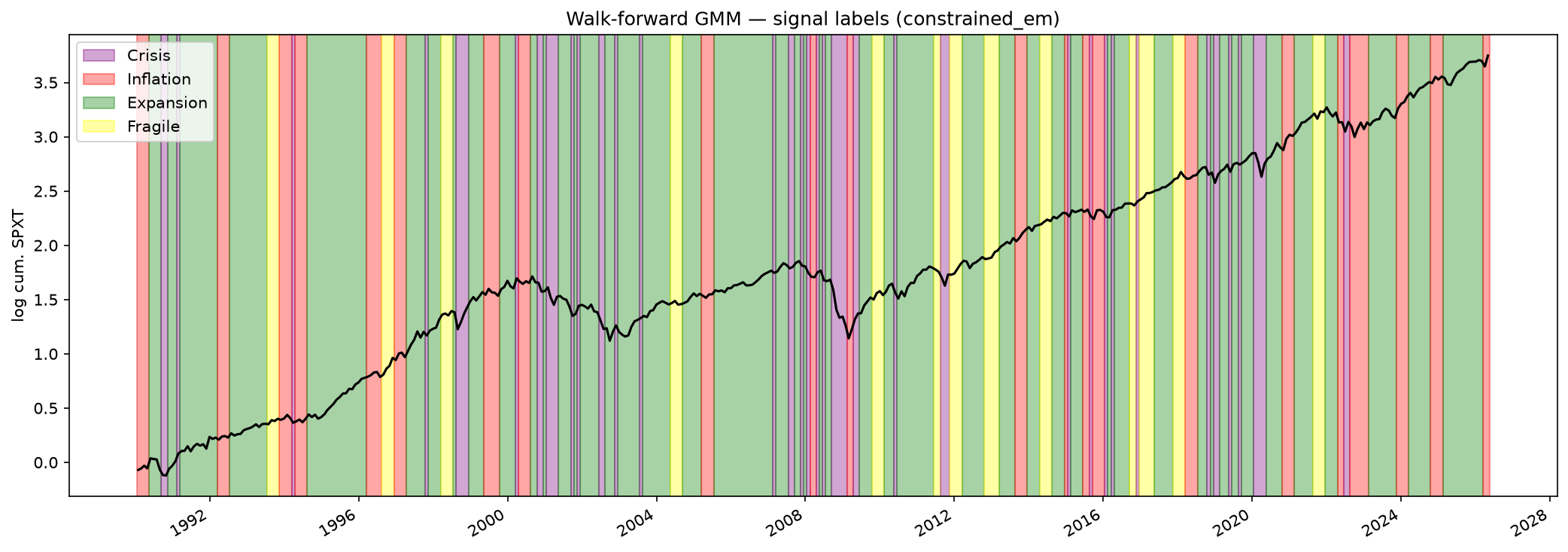}
\caption{Walk-forward classifications for the Gaussian mixture under constrained EM, January 1990 to April 2026. Colors and the equity return series are as in Figure~\ref{fig:oaci-timelines}.}
\label{fig:constrained-gmm}
\end{figure}

\paragraph{Constrained EM.}
Under constrained EM, the GMM retains all four regimes, assigning 251 of the 436 months (57.6\%) to Expansion alongside 56 to Crisis, 79 to Inflation, and 50 to Fragile (Figure~\ref{fig:constrained-gmm}). In the Student-$t$ models, the fit concentrates heavily into Expansion, which takes 75.5\% of months in the SMM (329) and 79.1\% in the SHMM (345), where Expansion spells average almost a year (Appendix~\ref{app:additional-timelines}). Fragile nearly disappears as a result, appearing in only eight SMM months (1.8\%) and four SHMM months (0.9\%).\\
Under Student-$t$ emissions, the scale bounds on equities, the VIX, and credit spreads imply a conditional standard deviation of at least $\sqrt{13.5}\approx 3.67$ standardized units in Crisis and at most $\sqrt{3.375}\approx 1.84$ in Expansion. Because Expansion can concentrate more density near its mean while Crisis must spread mass more widely on those coordinates, these restrictions may contribute to the concentration of months in Expansion (Appendix~\ref{app:constrained-em-details}).

\subsection{Comparison of identification rules}

The clearest difference between the semi-supervised procedure and constrained EM concerns which months they assign to regimes outside Expansion. With the same smoothing rule, the share of SMM months classified as Crisis rises from 3.2\% under constrained EM to 27.1\% under OACI-EM. In the SHMM, constrained EM assigns only four months to Fragile, too few to characterize that regime with much confidence. OACI-EM assigns it 32 months, none of which falls in an NBER recession, and industrial production grew by 2.9\% year on year on average during those months. Among the 88 non-anchor months classified as Crisis by OACI-EM, 14 fall in a recession (15.9\%) and industrial production growth averages 0.3\%. The distinction between Crisis and Fragile therefore has a recognizable counterpart in economic activity even after setting aside the pre-classified anchor months.\\
Constrained EM also concentrates its highest recession frequency in Crisis at 25.0\%, though over a much smaller set of months. Under Hungarian matching, Fragile behaves more like a recession state than Crisis itself, with 17.1\% of its months in recession and industrial production contracting by 0.8\% year on year on average. Average CPI inflation splits the methods in the same way: it is highest in the Inflation regime under both OACI-EM (3.2\%, compared with 2.8\% in Crisis) and constrained EM (3.8\%), yet higher in Crisis (3.1\%) than in Inflation (2.5\%) under Hungarian matching.\\
Taken together, the SHMM results give the clearest economic interpretation to the semi-supervised classifications. All four regimes recur, Crisis has the weakest industrial production growth and highest recession incidence, and Inflation has the highest average CPI inflation.

\section{Portfolio backtests}

The portfolio tests ask whether the economic distinctions recovered by the models help an investor decide what to hold in the following month. We begin with hand-allocated strategies that clearly define each regime's holdings. We then estimate four regime portfolios from historical returns and give every identification method the same set to trade, so that performance differences reflect only the regime scores and the trades they generate. In the final experiment, we hold the regime signal fixed and examine whether updating the four portfolios each month improves on weights estimated before the backtest.\\
Let $w_{k,t}$ denote the vector of asset weights assigned to regime $k$ at the end of month $t$. The smoothed score $\tilde g_t(k)$ determines the fraction of wealth allocated to that portfolio, giving the target weights for month $t+1$ as
$$
w_t=\sum_{k=0}^{K-1}\tilde g_t(k)\,w_{k,t}.
$$
For a portfolio specified in advance or estimated once, $w_{k,t}=w_k$ throughout the backtest, although the investor's holdings change as the regime scores evolve. We also consider an allocation that invests entirely in the portfolio selected by the discrete trading signal (Appendix~\ref{app:hard-signals}).\\
The main backtest begins in cash in January 1990 and covers 435 invested months from February 1990 through April 2026. Both strategies and benchmarks earn zero interest on cash and pay five basis points per dollar of risky assets bought or sold, with net performance summarized by compound annual growth (CAGR), the annualized Sharpe ratio\footnote{We calculate the annualized Sharpe ratio as $\sqrt{12}\,\bar r/s(r)$, where $\bar r$ and $s(r)$ are the sample mean and standard deviation of net monthly returns and cash earns zero interest.} ($r_f=0$), and maximum drawdown. Wealth curves compound an initial \$1 at net returns, while drawdowns measure declines from the running peak and are reported as positive loss magnitudes in the text and tables (Appendix~\ref{app:backtest-metrics}).

\subsection{Benchmarks and rules-based allocations}
\label{sec:handcrafted}

We evaluate five rules across the twelve combinations of model family and identification method. Two rules illustrate why the choice of identification method can matter more for some portfolios than for others. The bond-floor rule uses the Expansion score to set equity exposure, while the all-weather rule also uses the Inflation score to choose between commodities and Treasuries. The remaining three rules, risk-on/risk-off, safe haven, and cash shelter, vary the defensive allocation between Treasuries and cash (Appendix~\ref{app:handcrafted-strategies}). We compare the strategies with buy-and-hold US equities, represented by the S\&P~500 total return index (B0), and with an equal-weight portfolio of US equities, Treasuries, and commodities (B1). Over the evaluation period, equities compounded at 11.07\% with a maximum drawdown of 50.95\%, while equal weighting of the three assets returned 7.01\% per year with a maximum drawdown of 33.15\%. Their annualized Sharpe ratios were respectively 0.79 and 0.91.

\paragraph{Bond floor (low risk).} 
The Expansion portfolio holds 70\% in equities and 30\% in Treasuries, while each of the other three regime portfolios holds Treasuries exclusively. Weighting these portfolios by the smoothed scores gives
$$
w_t^{\mathrm{eq}}=0.7\,\tilde g_t(\mathrm{Expansion}),
\qquad
w_t^{\mathrm{bonds}}=1-w_t^{\mathrm{eq}}.
$$
The investor therefore retains at least 30\% in Treasuries, and a reassignment among Crisis, Inflation, and Fragile has no effect on holdings as long as the Expansion score stays the same. Under this rule, the semi-supervised Student-$t$ HMM (SHMM) had the highest sample Sharpe ratio of the twelve specifications, at 1.29, with annual compound growth of 7.44\% and a maximum drawdown of 18.91\%. The Gaussian mixture with Hungarian matching had a slightly smaller drawdown of 18.50\%, but compounded at 6.79\% with a Sharpe ratio of 1.20. The Student-$t$ mixture (SMM) with constrained EM grew faster at 8.14\% per year, with a lower Sharpe ratio of 1.12 and a drawdown of 19.90\%.

\begin{figure}[H]
\centering
\articlegraphic[width=0.8\textwidth]{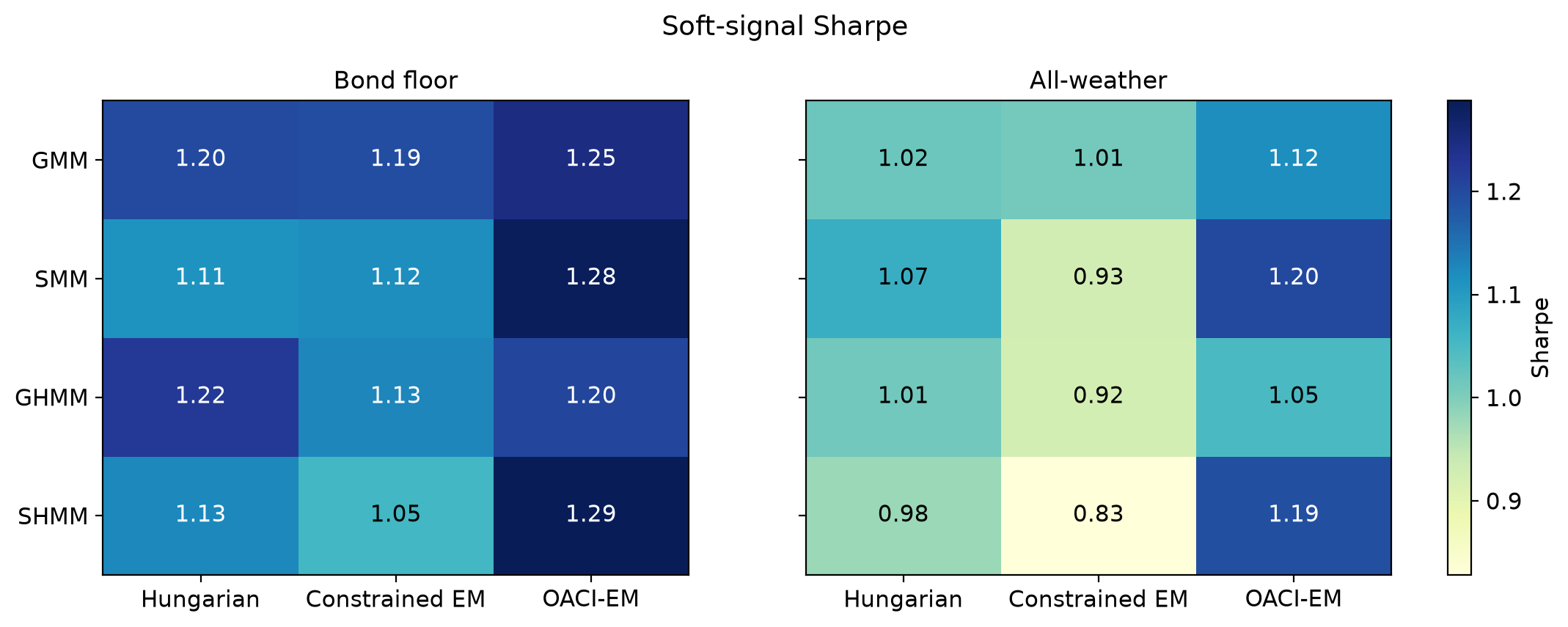}
\caption{Annualized Sharpe ratios for the score-weighted bond-floor and all-weather allocations, comparing the three identification methods within each model family.}
\label{fig:rules-based-sharpe}
\end{figure}

\begin{figure}[H]
\centering
\articlegraphic[width=\textwidth]{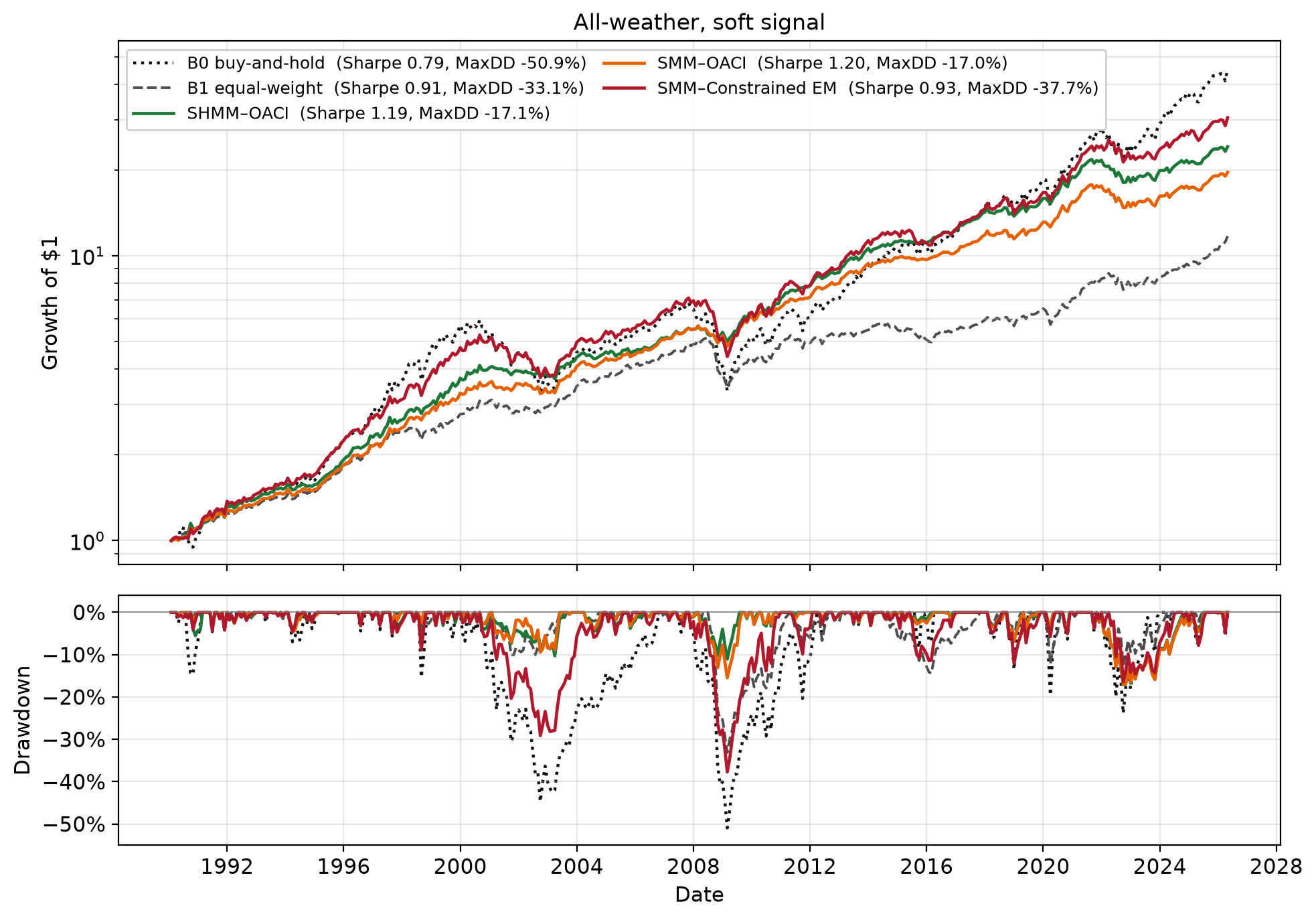}
\caption{Medium-risk all-weather rule under blended allocation. The panels
show growth of \$1 on a log scale and drawdown for B0, B1, SHMM-OACI,
SMM-OACI, and SMM-Constrained EM.}
\label{fig:all-weather-path}
\end{figure}

\paragraph{All-weather allocation (medium risk).}
The all-weather rule gives Inflation a distinct investment role by assigning equities to Expansion, commodities to Inflation, and Treasuries to Crisis and Fragile. The resulting asset weights are
$$
\begin{aligned}
w_t^{\mathrm{eq}}    &= \tilde g_t(\mathrm{Expansion}),\\
w_t^{\mathrm{bonds}} &= \tilde g_t(\mathrm{Crisis})+\tilde g_t(\mathrm{Fragile}),\\
w_t^{\mathrm{comm}}  &= \tilde g_t(\mathrm{Inflation}).
\end{aligned}
$$
Here a shift in the regime score from Crisis to Inflation moves capital from Treasuries to commodities even if the equity allocation stays constant. With this rule held fixed, semi-supervised identification gave a higher sample Sharpe ratio than either benchmark identification method in every model family (Figure~\ref{fig:rules-based-sharpe}). The difference was particularly pronounced for the Student-$t$ models. The SMM Sharpe ratio rose from 0.93 under constrained EM to 1.20 under OACI, and the SHMM Sharpe ratio rose from 0.83 to 1.19.\\
The two anchored strategies experienced similar drawdowns of about 17\%, with annual compound growth of 8.54\% for the SMM and 9.16\% for the SHMM. The constrained SMM achieved higher growth of 9.87\%, but its maximum drawdown reached 37.72\%. For the anchored SHMM, comparison with the passive benchmarks gives a sense of the trade-off. Its annual return fell short of buy-and-hold equities by 1.91 percentage points, while its maximum drawdown was 17.08\% against 50.95\% for equities. Against equal weighting of the same three assets, the strategy both compounded faster (9.16\% versus 7.01\%) and nearly halved the maximum drawdown (17.08\% versus 33.15\%). Part of the protection relative to equities comes from the ability to hold Treasuries and commodities. The equal-weight comparison holds the asset universe fixed, although differences in average exposure as well as changes in exposure over time contribute to the result.\\
Using the full score vector also mattered for these comparisons. Across the five rules and twelve specifications, score-weighted allocation had a higher sample Sharpe ratio than selecting a single regime portfolio in all 60 cases. For the anchored SHMM all-weather strategy, the corresponding Sharpe ratios were 1.19 and 0.81 (Appendix~\ref{app:hard-signals}). Figure~\ref{fig:all-weather-path} shows the all-weather wealth and drawdown paths. The corresponding bond-floor paths appear in Appendix~\ref{app:handcrafted-strategies} (Figure~\ref{fig:bond-floor-path}).

\subsection{Allocation with fixed regime portfolios}
\label{sec:data-driven}

\begin{figure}[H]
\centering
\articlegraphic[width=\textwidth]{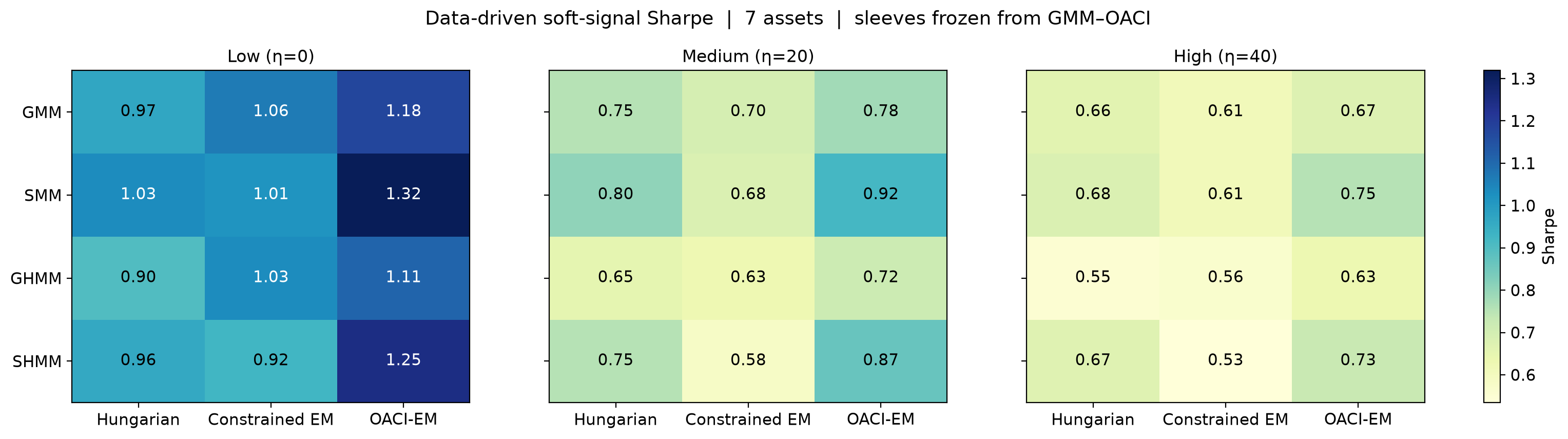}
\caption{Annualized Sharpe ratios for the seven-asset portfolios. The panels correspond to growth weights $\eta=0,20,40$. Within each panel, all twelve specifications use the same four regime portfolios estimated before January 1990.}
\label{fig:datadriven-heatmap-by-risk}
\end{figure}

We next allow the historical returns associated with each regime to determine its portfolio. The investment universe consists of US equities (\texttt{SPXT}), US Treasuries (\texttt{LUATTRUU}), US high-yield credit (\texttt{LF98TRUU}), commodities (\texttt{BCOMTR}), the US dollar (\texttt{DXY}), emerging-market equities (\texttt{MXEF}), and the SG Trend Index (\texttt{NEIXCTAT}). We estimate the four portfolios once, using data from March 1971 through December 1989, and keep their asset weights fixed throughout the evaluation. The seven-asset benchmark (B2) rebalances to equal weights and pays the same transaction costs as the strategies. Dollar exposure is measured using DXY index returns, which exclude funding, carry, and roll costs.\\
To partition the sample by regime, we fit a static Gaussian mixture with OACI-EM to the twelve standardized indicators (with September 1974 as the sole eligible anchor, floored at a 0.90 Inflation responsibility) and assign each usable month to the state with the largest fitted responsibility in \eqref{eq:gmm-estep}. This splits the 190 months into 17 Crisis, 28 Inflation, 127 Expansion, and 18 Fragile observations, leaving the covariance estimates for the smaller regimes sensitive to individual months.\\
Let $\bar r_k$ and $\widehat\Sigma_k$ denote the sample mean vector and covariance matrix of monthly asset returns over the $n_k$ months assigned to regime $k$, and let $\bar r$ denote the vector of mean returns across all usable months. We shrink each regime mean toward $\bar r$ with a weight equivalent to 24 observations and add a small diagonal term to the covariance matrix. The annualized estimates are
$$
\mu_{k,a}=12\,\frac{n_k\bar r_k+24\bar r}{n_k+24},
\qquad
\Sigma_{k,a}=12\bigl(\widehat\Sigma_k+10^{-5}I\bigr).
$$
For each regime, we choose a fully invested, long-only portfolio by solving
$$
w_k^\star(\eta)
\in
\operatorname*{arg\,max}_{w\geq0,\,\mathbf{1}^{\top}w=1}
\left\{
\frac{w^\top\mu_{k,a}}{\sqrt{w^\top\Sigma_{k,a}w}}
+
\eta\left(w^\top\mu_{k,a}-\tfrac12 w^\top\Sigma_{k,a}w\right)
\right\}.
$$
Setting $\eta=0$ (low growth) maximizes the estimated Sharpe ratio, while the medium- and high-growth settings ($\eta=20$ and $\eta=40$) place progressively more weight on approximate expected log growth. We solve each optimization by sequential least squares programming, using multiple starting points whenever $\eta>0$. Since the returns used to estimate each portfolio come from the months classified in that regime, applying its weights in the following month relies on some persistence in those return characteristics.\\
At each growth setting $\eta$, every model and identification method trades the same four portfolios, so specification $m$ sets its month-end target weights to
$$
w_t^{(m,\eta)}
=
\sum_{k=0}^{K-1}\tilde g_t^{(m)}(k)\,w_k^\star(\eta).
$$
Holding the four portfolios fixed yields the most consistent advantage for semi-supervised identification. Within each of the four model families, OACI produced a higher sample Sharpe ratio than both Hungarian matching and constrained EM at all three growth settings (Figure~\ref{fig:datadriven-heatmap-by-risk}). For portfolios restricted to equities, Treasuries, and commodities, OACI-EM also achieved the highest sample Sharpe ratio within each model family at all three growth settings (Appendix~\ref{app:data-driven-sleeves}, Table~\ref{tab:dd-sharpe-3}). Because the four portfolios are estimated from the pre-1990 Gaussian OACI partition, these rankings may favor signals that attach similar meanings to the labels. Relative to passive allocation, however, only the low-growth setting in Table~\ref{tab:seven-asset-performance} beats the equal-weight Sharpe ratio for either anchored Student-$t$ model.

\begin{table}[ht]
\centering
\small
\setlength{\tabcolsep}{4pt}
\begin{tabular}{@{}llrrr@{}}
\toprule
Portfolio & Growth setting & CAGR (\%) & Max. drawdown (\%) &
\shortstack{Annualized Sharpe\\($r_f=0$)} \\
\midrule
US equities (B0) & & 11.07 & 50.95 & 0.79 \\
Seven assets, equal weight (B2) & & 6.01 & 24.49 & 0.93 \\
\addlinespace
SMM with OACI  & Low    & 5.71 &  8.51 & 1.32 \\
SHMM with OACI & Low    & 5.94 &  7.26 & 1.25 \\
SMM with OACI  & Medium & 7.91 & 20.23 & 0.92 \\
SHMM with OACI & Medium & 8.10 & 29.40 & 0.87 \\
SMM with OACI  & High   & 7.31 & 28.98 & 0.75 \\
SHMM with OACI & High   & 7.74 & 35.62 & 0.73 \\
\bottomrule
\end{tabular}
\caption{Fixed seven-asset regime portfolios and passive benchmarks, January 1990 to April 2026. Performance is net of trading costs, and maximum drawdown is expressed as a loss magnitude.}
\label{tab:seven-asset-performance}
\end{table}

\begin{figure}[H]
\centering
\articlegraphic[width=\textwidth]{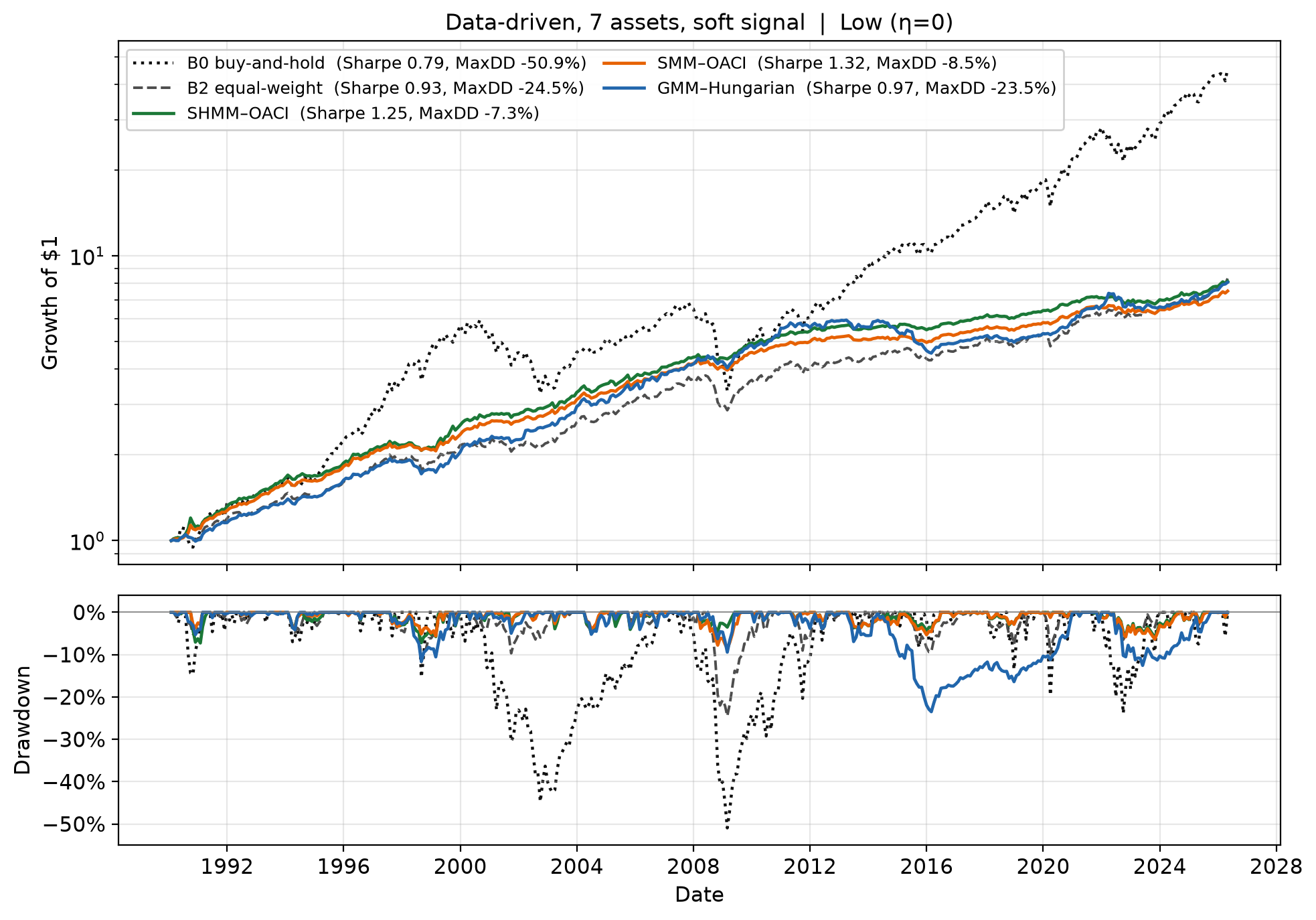}
\caption{Growth of \$1 (log scale) and drawdown for the score-weighted low-growth seven-asset portfolios ($\eta=0$) alongside passive benchmarks B0 and B2.}
\label{fig:datadriven-eight-path-low}
\end{figure}

\paragraph{Low growth emphasis ($\eta=0$).}
Even in Expansion, the estimated portfolio allocates approximately 46\% to Treasuries and 20\% to high-yield credit. The Crisis portfolio holds about 55\% in Treasuries and 37\% in the dollar, while the Inflation portfolio invests entirely in commodities. With these relatively defensive holdings, the anchored SHMM compounded at 5.94\%, close to the 6.01\% earned by equal weighting of the seven assets, while its maximum drawdown fell from the benchmark's 24.49\% to 7.26\% (Figure~\ref{fig:datadriven-eight-path-low}). The anchored SMM had the highest sample Sharpe ratio, at 1.32, although it compounded somewhat more slowly and experienced a slightly deeper drawdown than the anchored SHMM.

\paragraph{Medium and high growth emphasis ($\eta=20,40$).}
At the medium setting, Expansion holds 32\% in US equities and 68\% in emerging-markets, while Fragile retains a 54\% Treasury allocation. Both anchored Student-$t$ strategies compounded faster than the seven-asset benchmark, reaching 7.91\% for the SMM and 8.10\% for the SHMM. Regarding drawdowns, the SMM lost 20.23\% from peak to trough, less than the benchmark's 24.49\%, whereas the SHMM lost 29.40\%.\\
At the high setting, both Expansion and Fragile invest entirely in equities. Their respective allocations to US and emerging-market equities are 21\% and 79\% in Expansion, and 43\% and 57\% in Fragile. Whether separating two regimes changes the portfolio much depends on how far apart their holdings are. Moving ten percentage points of score from Expansion to Fragile shifts only 2.2 percentage points of the portfolio from emerging-market to US equities, while the same shift from Crisis to Inflation moves a full ten percentage points from the dollar into commodities. For both anchored Student-$t$ models, compound growth peaked at the medium setting, while the high setting produced the deepest drawdown and the lowest Sharpe ratio. The corresponding wealth paths appear in Appendix~\ref{app:data-driven-sleeves} (Figures~\ref{fig:datadriven-eight-path-medium} and~\ref{fig:datadriven-eight-path-high}).

\begin{figure}[H]
\centering
\articlegraphic[width=\textwidth]{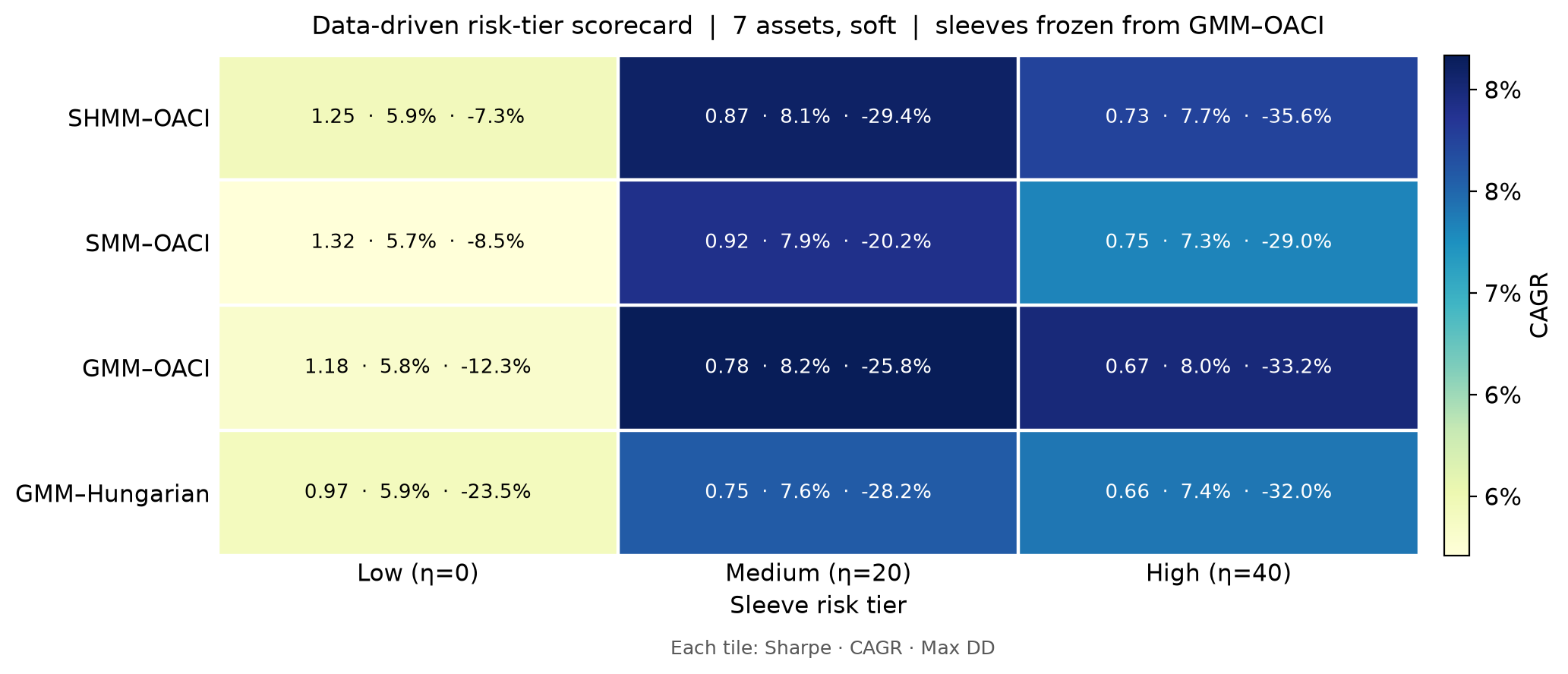}
\caption{Seven-asset performance across growth settings for selected specifications. Each cell gives the annualized Sharpe ratio, CAGR, and maximum drawdown, with color indicating CAGR.}
\end{figure}

\subsection{Monthly portfolio updates using same-month or subsequent returns}

The fixed portfolios use only returns observed through December 1989. We now retain the smoothed scores from the semi-supervised SHMM and allow the asset weights within each regime portfolio to evolve as further returns become available. At every month-end, the two updating strategies use the same expanding history of seven-asset returns, beginning with 60 observations. For a return realized in month $u$, the same-month rule uses $\tilde g_u(k)$ to estimate how the assets performed during the conditions identified in that month, whereas the next-month rule uses $\tilde g_{u-1}(k)$ to estimate how they performed in the month following those conditions.\\
For example, at the end of January both strategies can use the return just observed in January. The same-month rule pairs it with January's regime scores and the next-month rule pairs it with December's. Each strategy uses these historical pairs to estimate four fully invested, long-only portfolios that maximize the Sharpe ratio subject to a 40\% cap on any one asset. Both then combine their updated portfolios using January's scores to determine the holdings for February (see Appendix~\ref{app:allocation-controls} for the estimators).

The two monthly updates compounded at almost the same rate, 5.06\% per year for the same-month rule and 5.05\% for the next-month rule, although the same-month rule had a smaller maximum drawdown (4.80\% versus 6.77\%) and a higher Sharpe ratio (1.56 versus 1.48). The paired 95\% bootstrap interval for the Sharpe difference, measured as next-month minus same-month, was $[-0.25,0.05]$, so the data do not establish an advantage for either pairing. Over the same period, the fixed portfolios compounded at 5.17\% with a maximum drawdown of 7.19\%. Their comparison with the monthly updates also reflects differences in the estimation sample, the way historical returns are assigned to regimes, the covariance regularization, and the asset limits.

\begin{figure}[H]
\centering
\articlegraphic[width=\textwidth]{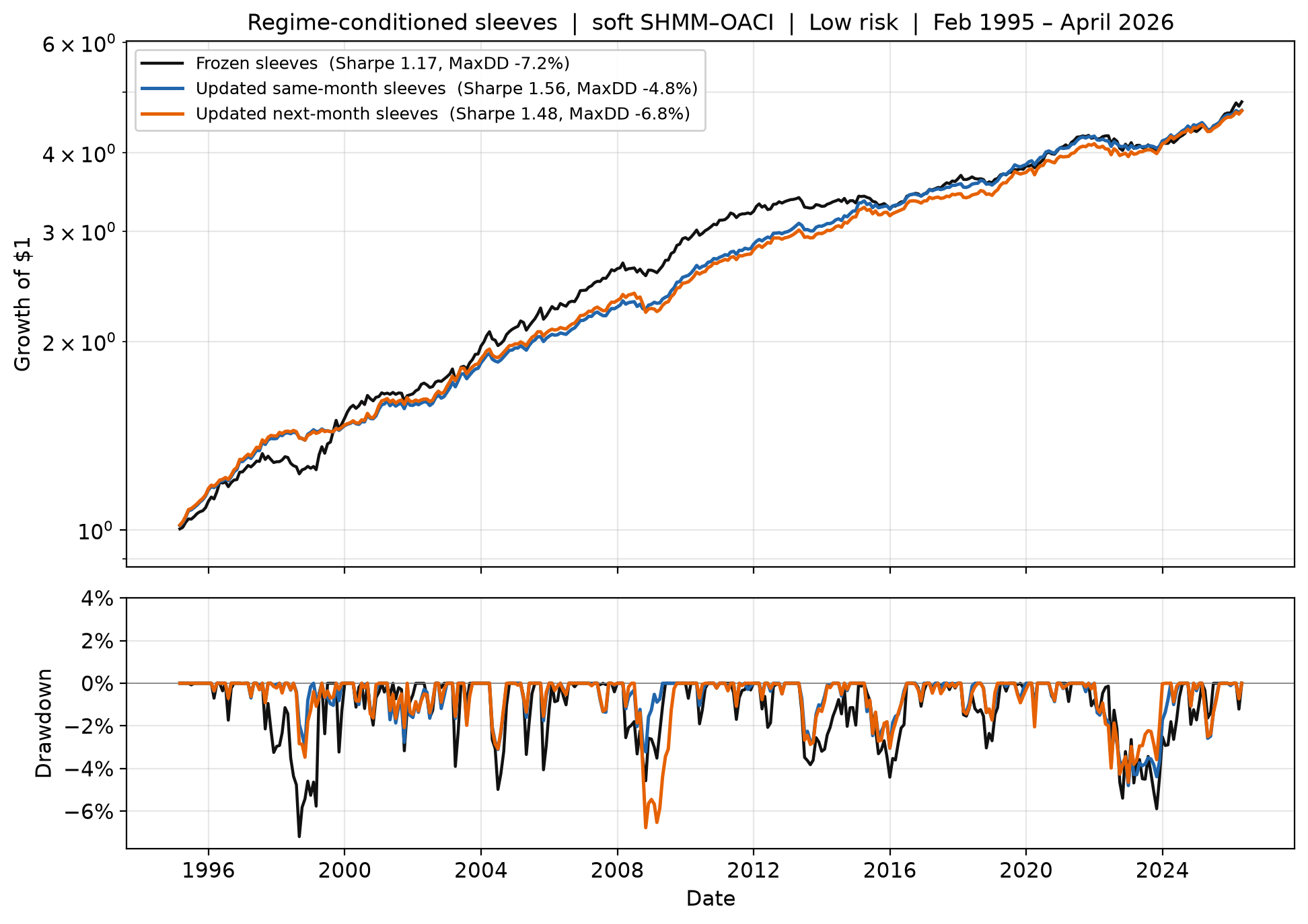}
\caption{Growth of \$1 and drawdown for the fixed low-growth portfolios and the two monthly updates. All three strategies start in cash in January 1995, use the same anchored SHMM scores, and invest from February 1995 through April 2026.}
\end{figure}

\subsection{Discussion}

The three-asset all-weather strategy shows most clearly how semi-supervised identification benefits an investor. Holding the model family and allocation rule fixed, replacing constrained EM with OACI-EM raises the Student-$t$ HMM Sharpe ratio from 0.83 to 1.19. Over the evaluation period, the anchored strategy retains more than four fifths of buy-and-hold US equity compound growth (9.16\% versus 11.07\%) while cutting maximum drawdown by roughly two thirds, from 50.95\% to 17.08\%. Compared with equal weighting of the same three assets, it compounds faster (9.16\% versus 7.01\%), nearly halves the maximum drawdown (17.08\% versus 33.15\%), and lifts the Sharpe ratio from 0.91 to 1.19.\\
In the seven-asset backtests, OACI-EM achieves the highest sample Sharpe ratio within every model family across all three growth settings, while the growth parameter governs the return-to-drawdown trade-off through the Expansion weights. At low growth, where Expansion still holds roughly 46\% in Treasuries and 20\% in high-yield credit, the anchored SHMM nearly matches seven-asset equal-weight growth (5.94\% versus 6.01\%) while compressing maximum drawdown from 24.49\% to 7.26\%. Moving Expansion fully into equities at the medium setting lifts annual growth to 8.10\%, nearly three quarters of US equity compounding, though drawdown widens to 29.40\%, and emphasizing growth further lowers realized return while deepening losses. These holdings also determine which regime shifts move capital, as transferring probability from Crisis to Inflation leaves the bond floor unchanged but shifts Treasuries into commodities under the all-weather rule and the dollar into commodities at high growth. Even when equal weighting controls for the asset universe, both average defensive exposure and trade timing contribute to this downside protection.\\
Finally, comparing the monthly updates with an investor who pools the same seven-asset return history to estimate a single maximum-Sharpe portfolio under the same 40\% asset cap separates allocation from weight estimation. The next-month rule beats the pooled portfolio by 0.09 in sample Sharpe, with a paired 95\% bootstrap interval of $[-0.02,0.22]$ (Appendix~\ref{app:allocation-controls}). Semi-supervised identification therefore shows its clearest advantage when allocating across a fixed set of regime portfolios, while its value for re-estimating portfolio weights monthly remains uncertain over this sample.

\section{Conclusion and Further Perspectives}

Semi-supervised identification gave the clearest economic interpretation of the fitted regimes and the most consistent improvement in risk-adjusted performance when the models traded a common set of regime portfolios. One reason it may help is that the templates and the historical labels answer different questions. A template describes the pattern we expect across markets, while a labeled episode gives the model an observed example of how those markets behaved together. In the anchored Student-$t$ hidden Markov model, the distinction between Crisis and Fragile remained economically meaningful outside the anchor months. Recessions were most frequent and industrial production growth weakest in Crisis, while the months classified as Fragile showed signs of financial stress alongside continued growth in production.\\
For an investor, the value of that distinction depends on the holdings it changes. The anchored all-weather strategy retained much of the compound growth of US equities with a substantially shallower drawdown, and it both compounded faster and experienced a smaller drawdown than equal weighting of the same assets. The clearest comparison of identification methods came from the fixed regime portfolios, where each model family achieved its highest sample Sharpe ratio under semi-supervision at every growth setting. There was less evidence that regime scores improved the estimation of portfolio weights themselves. In the monthly updates, the advantage over estimating a single portfolio from pooled returns remained uncertain.\\
Delaying each anchor preserves chronology, but the episodes were selected retrospectively, leaving open how a pre-specified calendar would perform. Indeed, the anchored model initially classified the 2022 inflation episodes as Crisis. Attribution also matters for the fixed portfolios, which come from an anchored partition and owe part of their drawdown protection to defensive holdings.\\
A direct test would set aside entire episodes before choosing the anchors and compare fits with and without dated labels while keeping the templates, parameter restrictions, and scoring rule fixed. Scoring the same observations under successive fits would then show whether the labels retain their meaning as the estimation window expands. For the portfolio comparison, a benchmark rebalanced to the strategy's average holdings through the preceding month would help distinguish the contribution of changing exposures from that of a persistently defensive allocation.\\
Inflation and financial stress can hit at the same time, widening credit spreads and pushing down equities while our current model assigns each month to a single latent state. Modeling inflation pressure and financial stress with separate latent states, each keeping its own historical anchors while letting their joint configuration vary, would let us test across economies with different inflation histories whether financial stress affects future output growth and government-bond protection differently when inflation is high rather than subdued.

\bibliographystyle{elsarticle-harv}
\bibliography{references}

\newpage
\appendix

\section{Data provenance}
\label{app:data-provenance}

Table~\ref{tab:factor-provenance} records, for each column of the core panel, the Bloomberg series, the native start (first month of the transformed return or difference, not the first date in the file), and the substitute used before that month. The panel contains 662 monthly observations from March 1971 through
April 2026. We exclude the partial May observation. Pre-inception
index histories and proxy returns are historical reconstructions.\\
The constant-duration ten-year Treasury stand-in (with yield $y_{10,t}$ in percent), also used inside the trend proxy, is
\begin{equation}
r^{\mathrm{proxy}}_t = -7 \cdot \Delta y_{10,t}/100 + y_{10,t-1}/1200,
\label{eq:bond-proxy}
\end{equation}
and the pre-2000 trend stand-in is
\begin{equation}
\begin{aligned}
r^{\mathrm{trend}}_t
&= \tfrac12 \, s^{\mathrm{com}}_{t-1} \, r^{\mathrm{BCOM}}_t
+ \tfrac12 \, s^{y}_{t-1} \, r^{\mathrm{proxy}}_t, \\
s^{\mathrm{com}}_{t-1} &= 2\,\mathbf{1}\!\left\{R^{\mathrm{BCOM}}_{t-1,12}>0\right\}-1, \\
s^{y}_{t-1} &= 2\,\mathbf{1}\!\left\{y_{10,t-1}-y_{10,t-13}>0\right\}-1.
\end{aligned}
\label{eq:trend-proxy}
\end{equation}
The commodity leg uses the trailing twelve-month commodity return, and the Treasury leg uses the change over the previous year in the ten-year yield. A positive signal gives a long position in the following month. A zero, negative, or unavailable signal gives a short position in the backfilled sample. The yield-sign rule can produce a position opposite to the one indicated by trailing Treasury total-return momentum.

\begin{table}[!ht]
\centering
\caption{Observed history and pre-inception substitutes for the twelve-indicator panel.}
\label{tab:factor-provenance}
\footnotesize
\setlength{\tabcolsep}{3.5pt}
\renewcommand{\arraystretch}{1.08}
\begin{tabular}{@{}p{0.15\textwidth}p{0.20\textwidth}p{0.10\textwidth}p{0.50\textwidth}@{}}
\toprule
Column & Bloomberg series & Native start & Substitute before the native start \\
\midrule
\texttt{SPXT}
  & S\&P 500 total return
  & 1989-10
  & Price return of \texttt{SPX} (no dividends in identification sample; 1971-1989 portfolio estimation adds 4\% annualized dividend yield). \\
\texttt{LUATTRUU}
  & US Treasury total return
  & 1973-02
  & Eq.~\eqref{eq:bond-proxy}. \\
\texttt{YC\_10Y\_3M}
  & \texttt{USGG10YR} minus \texttt{USGG3M}
  & 1971-03
  & None (first difference of yield spread, in percentage points). \\
\texttt{LF98TRUU}
  & US Corp.\ High Yield TR
  & 1983-08
  & \texttt{LUACTRUU} (US Corp.\ IG total return) from 1973-02; Eq.~\eqref{eq:bond-proxy} earlier. \\
\texttt{LUACOAS}
  & US Aggregate Credit OAS
  & 1989-07
  & First diff.\ of Moody's Baa$-$Aaa (\texttt{MOODCBAA}$-$\texttt{MOODCAAA}), in percent (yield spread, not OAS). \\
\texttt{VIX}
  & CBOE VIX
  & 1990-02
  & Percent change of month-end 30-day realized vol from daily \texttt{SPX} ($\sigma\sqrt{252}\times 100$); VXO unused. \\
\texttt{BCOMTR}
  & Bloomberg Commodity TR
  & 1971-03
  & Vendor series from January 1971. Its historical extension uses 1998 commodity target weights, rescaled over eligible contracts. \\
\texttt{BCIT1T}
  & US Gov.\ Inflation-Linked, 1+ yr TR
  & 1997-03
  & Latest published $\Delta$ YoY CPI (pp) via historical BLS release calendar and revised Bloomberg CPI extract. From Mar.\ 1997: decimal monthly TIPS index return. \\
\texttt{DXY}
  & US Dollar Index
  & 1971-03
  & None in code; pull starts Jan.\ 1971 (includes pre-1973 history). \\
\texttt{MXEF}
  & MSCI Emerging Markets (USD price index)
  & 1988-01
  & Monthly price return of MSCI World (\texttt{MXWO}, USD) through December 1987. The Emerging Markets index starts at 100 on 31 Dec.\ 1987. Both price indices exclude regular cash dividends. \\
\texttt{NEIXCTAT}
  & SG Trend Index
  & 2000-01
  & Eq.~\eqref{eq:trend-proxy}; rebased at 1{,}000 on 31 Dec.\ 1999. \\
\texttt{CU\_AU\_RATIO}
  & \texttt{HG1} copper \& \texttt{XAU} gold
  & 1989-01
  & Log change of \texttt{BCOMTR} minus log change of gold before copper is available. \\
\bottomrule
\end{tabular}
\end{table}

\paragraph{Macroeconomic series for Table~\ref{tab:shmm-occupancy}.}
The CPI column uses the workbook's year-on-year percentage change in the BLS all-items CPI for urban consumers in the US city average, not seasonally adjusted (\texttt{CPIAUCNS}).\footnote{US Bureau of Labor Statistics, historical CPI-U tables, \url{https://www.bls.gov/regions/mid-atlantic/data/consumerpriceindexhistorical_us_table.htm}.}
The IP column uses the Federal Reserve's seasonally adjusted total industrial production index (G.17 series \texttt{B50001}, corresponding to \texttt{INDPRO}), from the 17 July 2026 release.\footnote{Federal Reserve Board, G.17 historical data, \url{https://www.federalreserve.gov/releases/g17/20260717/ipdisk/table1_2.htm}.}
We calculate IP growth as the percentage change from the workbook's level twelve months earlier and average each growth series over its available non-anchor observations.

\paragraph{Release timing for the CPI proxy.}
Before March 1997, the inflation-linked coordinate \texttt{BCIT1T}
uses the change in year-on-year US CPI inflation most recently
published by the month-end decision date. Let $m(t)$ denote the
latest CPI reference month released by the US Bureau of Labor Statistics
on or before month-end $t$. The proxy is
\[
r^{\mathrm{CPI}}_t=\pi_{m(t)}-\pi_{m(t)-1},
\]
where $\pi_m$ is year-on-year inflation in percentage points.
A month without a release retains the previous available change.
When two releases occur in one month, the later reference month
determines the feature. For example, December 1995 CPI was released
on 1 February 1996 and was unavailable at the January decision.\footnote{US Bureau of Labor Statistics, historical CPI release dates, \url{https://www.bls.gov/bls/histreleasedates.pdf}.}\\
The release calendar determines availability, while CPI values come
from the revised Bloomberg historical extract. Historical revisions
can therefore affect the proxy. Beginning in March 1997 the coordinate
uses the observed monthly TIPS index return. The splice changes both
the economic quantity and its units. The release-dated series enters
standardization, the pre-1990 regime partition used to estimate
portfolio sleeves, and subsequent walk-forward estimation.

\paragraph{Standardization of monthly changes.}
Monthly changes $r_{t,j}$, $j=1,\ldots,d$, are constructed as percentage changes in index levels and VIX, first differences in yield and credit spreads, and the difference between copper and gold log returns. Historical proxies follow the conventions in Appendix~\ref{app:data-provenance}.\\
Each series is standardized using exponentially weighted moments with span $S=60$, decay $\alpha=2/(S+1)$, and a 36-observation minimum. Starting from $\widehat{\mu}_{1,j}=r_{1,j}$ and $V_{1,j}=0$, for $t\ge 2$,
\begin{equation}
\begin{aligned}
\widehat{\mu}_{t,j}
&= (1-\alpha)\widehat{\mu}_{t-1,j}+\alpha r_{t,j},\\[0.3em]
V_{t,j}
&= (1-\alpha)V_{t-1,j}
+\alpha(1-\alpha)\bigl(r_{t,j}-\widehat{\mu}_{t-1,j}\bigr)^2.
\end{aligned}
\end{equation}
The weights sum to one. Their squared sum satisfies $q_1=1$ and $q_t=(1-\alpha)^2q_{t-1}+\alpha^2$, giving the bias-adjusted standard deviation
\begin{equation}
\widehat{\sigma}_{t,j}
=\sqrt{\frac{V_{t,j}}{1-q_t}},
\qquad t\ge 2.
\end{equation}
Month $t$ is standardized using moments through $t-1$:
\begin{equation}
x_{t,j}
=
\frac{r_{t,j}-\widehat{\mu}_{t-1,j}}
{\max\{\widehat{\sigma}_{t-1,j},\,10^{-8}\}}.
\end{equation}
The denominator floor prevents division by zero, and each coordinate is clipped at $\pm4$. The first admissible observation is the 37th sample month. Under expanding-window estimation, earlier standardized rows remain unchanged as new months enter. A positive standardized value denotes a monthly change above its trailing average, even when the raw change is negative.

\section{Baseline model specifications}
\label{app:baseline-models}

All fits use at most 200 iterations with tolerance $10^{-4}$. The specifications below describe the models before economic identification.

\subsection{Mixture models}

Finite mixtures model observations independently, with no transition mechanism linking successive regimes \citep{mclachlan2000finite}. Persistence in their inferred signals can nevertheless arise from serial dependence in the observed indicators and subsequent temporal smoothing.

\paragraph{Gaussian mixture model (GMM).}
Let $\pi_k=\Pr(z_t=k)$, with $\pi_k\ge0$ and $\sum_{k=0}^{K-1}\pi_k=1$. Component $k$ has mean $\mu_k\in\mathbb{R}^d$ and diagonal covariance $\mathrm{diag}(\sigma_k^2)$, where $\sigma_k^2\in\mathbb{R}_+^d$. For $\Theta=\{\pi_k,\mu_k,\sigma_k^2\}_{k=0}^{K-1}$,
\begin{equation}
p(x_t;\Theta)
=
\sum_{k=0}^{K-1}\pi_k\,
\mathcal{N}\!\bigl(x_t;\mu_k,\mathrm{diag}(\sigma_k^2)\bigr).
\end{equation}
Estimation uses regularized EM updates \citep{dempster1977maximum}. The E-step computes posterior responsibilities,
\begin{equation}
\label{eq:gmm-estep}
\gamma_{t,k}
=
\frac{\pi_k\,\mathcal{N}\!\bigl(x_t;\mu_k,\mathrm{diag}(\sigma_k^2)\bigr)}
{\sum_{j=0}^{K-1}\pi_j\,\mathcal{N}\!\bigl(x_t;\mu_j,\mathrm{diag}(\sigma_j^2)\bigr)},
\end{equation}
and the M-step updates mixing weights, means, and variances in closed form (Appendix~\ref{app:em-updates}). We use a symmetric Dirichlet prior with $\alpha_0=1.5$ on the mixing weights and a variance floor of $10^{-4}$. Ten $K$-means-initialized fits are compared, retaining the highest sample likelihood. Responsibilities give soft assignments, while $\hat{z}_t=\arg\max_k\gamma_{t,k}$ gives hard assignments before temporal smoothing.

\paragraph{Student-$t$ mixture model (SMM).}
The SMM replaces Gaussian components with multivariate Student-$t$ densities to accommodate heavy tails \citep{peel2000robust}:
\begin{equation}
p(x_t;\Theta)
=
\sum_{k=0}^{K-1}\pi_k\,
t_{\nu}\!\bigl(x_t;\mu_k,\mathrm{diag}(\sigma_k^2)\bigr),
\end{equation}
with fixed $\nu=6$. Here $\mathrm{diag}(\sigma_k^2)$ is the scale matrix, whose corresponding covariance is $\nu/(\nu-2)$ times that matrix. Estimation uses Expectation Conditional Maximization (ECM) \citep{meng1993maximum}, with the GMM's mixing-weight prior, numerical floor, and initialization settings. The emission updates incorporate latent weights $u_{t,k}=(\nu+d)/(\nu+\delta_{t,k})$, where $\delta_{t,k}$ is squared Mahalanobis distance computed using the component scale matrix. Observations farther from a component receive less weight in its mean and scale updates. Appendix~\ref{app:em-updates} gives the distinct normalization terms for these updates.

\subsection{Hidden Markov models}

HMMs retain conditionally independent emissions given the state sequence but link states through a first-order Markov chain \citep{hamilton1989new, ryden1998stylized}. The transition matrix $A=(a_{jk})$, with $a_{jk}=\Pr(z_t=k\mid z_{t-1}=j)$, models persistence and switching. Here $\pi$ denotes the initial state distribution.

\paragraph{Gaussian HMM (GHMM).}
The emissions are diagonal Gaussian,
\begin{equation}
x_t\mid z_t=k
\sim
\mathcal{N}\!\bigl(\mu_k,\mathrm{diag}(\sigma_k^2)\bigr),
\end{equation}
with parameters $\Theta=\{\pi,A,\{\mu_k,\sigma_k^2\}_{k=0}^{K-1}\}$. Baum-Welch estimation combines EM with forward-backward recursions, using smoothed state probabilities $\Pr(z_t=k\mid x_{1:T})$ and pairwise transition marginals within each training sample. Out-of-sample scoring uses the causal filtered posterior $\Pr(z_t=k\mid x_{1:t})$ (Section~\ref{sec:walk-forward}). We floor emission variances at $10^{-3}$ and retain the highest-likelihood fit from 8 static or 5 walk-forward restarts.\\
Let $N_{jk}$ denote the expected number of $j$-to-$k$ transitions
in the training sample. Both HMMs use a row-wise MAP update with these
counts and a shared Dirichlet prior. In the state order used during
estimation, the update and prior parameters are
\[
a_{jk}^{+}=
\frac{N_{jk}+\alpha_{jk}-1}
{\sum_{\ell=0}^{K-1}(N_{j\ell}+\alpha_{j\ell}-1)},
\qquad
(\alpha_{jk})=
\begin{pmatrix}
20 & 1 & 5 & 1 \\
1 & 20 & 4 & 1 \\
1 & 1 & 25 & 3 \\
5 & 1 & 2 & 15
\end{pmatrix}.
\]
The additional diagonal prior weight is set to zero.

\paragraph{Student-$t$ HMM (SHMM).}
The SHMM replaces Gaussian emissions with multivariate Student-$t$ densities,
\begin{equation}
x_t\mid z_t=k
\sim
t_{\nu}\!\bigl(\mu_k,\mathrm{diag}(\sigma_k^2)\bigr),
\end{equation}
with $\nu=6$ and the same scale-matrix interpretation as the SMM. ECM Baum-Welch combines forward-backward inference with latent scale weights that reduce the influence of distant observations on emission updates. The numerical floor, transition prior, and restart counts match the GHMM. Heavy-tailed emissions do not resolve the state-labeling ambiguity.

\section{EM and ECM Updates for Mixture Models}
\label{app:em-updates}

The M-step updates for the Gaussian mixture model (GMM) and the Student-$t$ mixture model (SMM) share a unified weighted-moment form. Let $f_k(x_t; \Theta^{(m)})$ denote the component density at iteration $m$, given by $\mathcal{N}\bigl(x_t; \mu_k^{(m)}, \mathrm{diag}(\sigma_k^{2,(m)})\bigr)$ for the GMM and $t_{\nu}\bigl(x_t; \mu_k^{(m)}, \mathrm{diag}(\sigma_k^{2,(m)})\bigr)$ for the SMM. In the E-step, the posterior responsibilities are
\begin{equation}
\gamma_{t,k}^{(m)} = \frac{\pi_k^{(m)} \, f_k(x_t; \Theta^{(m)})}{\sum_{l=0}^{K-1} \pi_l^{(m)} \, f_l(x_t; \Theta^{(m)})},
\end{equation}
and the observation-specific scale weights $u_{t,k}^{(m)}$ are defined as
\begin{equation}
\label{eq:app-mix-u}
u_{t,k}^{(m)} =
\begin{cases}
1, & \text{(GMM)}, \\[6pt]
\dfrac{\nu + d}{\nu + \sum_{j=1}^{d} \bigl(x_{t,j} - \mu_{k,j}^{(m)}\bigr)^2 / \sigma_{k,j}^{2,(m)}}, & \text{(SMM)}.
\end{cases}
\end{equation}\\
For Student-$t$ models with $\nu=6$, the parameters $\sigma_{k,j}^{2}$
are component scale parameters, not variances: the implied variance is
$\frac{\nu}{\nu-2}\sigma_{k,j}^{2}=1.5\,\sigma_{k,j}^{2}$. Template
volatility scores and Constrained-EM variance boxes are applied to these
scale parameters.\\
Writing $N_k = \sum_{t=1}^{T} \gamma_{t,k}^{(m)}$, the M-step (or CM-step) updates under a symmetric Dirichlet prior ($\alpha_0 = 1.5$) and variance floor $\varepsilon = 10^{-4}$ are
\begin{align}
\pi_k^{(m+1)} &= \frac{N_k + \alpha_0 - 1}{T + K(\alpha_0 - 1)}, \\[4pt]
\mu_{k,j}^{(m+1)} &= \frac{\sum_{t=1}^{T} \gamma_{t,k}^{(m)} u_{t,k}^{(m)} x_{t,j}}{\sum_{t=1}^{T} \gamma_{t,k}^{(m)} u_{t,k}^{(m)}}, \qquad j = 1, \ldots, d, \\[4pt]
\bigl(\sigma_{k,j}^{2}\bigr)^{(m+1)} &= \max\!\left( \frac{1}{N_k} \sum_{t=1}^{T} \gamma_{t,k}^{(m)} u_{t,k}^{(m)} \bigl(x_{t,j} - \mu_{k,j}^{(m+1)}\bigr)^2, \; \varepsilon \right), \qquad j = 1, \ldots, d.
\end{align}

\section{Hungarian Matching Implementation Details}
\label{app:hungarian-details}

\paragraph{Economic matching.}
In the compatibility score \eqref{eq:hungarian-score}, the weight
$w_j$ adjusts for the dispersion of feature $j$ across the $K$
fitted components:
\begin{equation}
w_j =
\left(\frac{1}{K}\sum_{i=0}^{K-1}\sigma_{i,j}^2\right)^{-1/2}.
\end{equation}
Here $\sigma_{i,j}^2$ is the fitted Gaussian variance or the diagonal
Student-$t$ scale entry. The calculation floors each entry at $10^{-12}$ before
averaging. This weighting reduces the influence of features with large
fitted dispersion.\\
The multiplier $b_{r,j}$ gives additional weight to feature $j$
when scoring economic regime $r$. Table~\ref{tab:app-hungarian-weights}
reports these multipliers.

\begin{center}
\begin{minipage}{\linewidth}
\centering
\refstepcounter{table}\label{tab:app-hungarian-weights}
\textbf{Table \thetable.} Economic matching weights.
\par\smallskip
\small
\begin{tabular}{@{}lrrrr@{}}
\toprule
& \multicolumn{4}{c}{Economic multiplier $b_{r,j}$} \\
\cmidrule(lr){2-5}
Feature & Crisis (0) & Inflation (1) & Expansion (2) & Fragile (3) \\
\midrule
\texttt{SPXT}          & 1 & 1    & 1.25 & 1.25 \\
\texttt{VIX}           & 1 & 1    & 1.5  & 1.5  \\
\texttt{LUACOAS}       & 1 & 1    & 1.5  & 1.5  \\
\texttt{LF98TRUU}      & 1 & 1    & 2    & 1    \\
\texttt{BCOMTR}        & 1 & 2    & 1    & 1    \\
\texttt{YC\_10Y\_3M}   & 1 & 1.5  & 1    & 1    \\
\texttt{LUATTRUU}      & 1 & 1    & 2    & 1    \\
\texttt{DXY}           & 1 & 1.25 & 1    & 1    \\
\texttt{MXEF}          & 1 & 1    & 1    & 1.5  \\
\texttt{CU\_AU\_RATIO} & 1 & 1    & 1    & 1    \\
\texttt{BCIT1T}        & 1 & 2    & 1    & 1    \\
\texttt{NEIXCTAT}      & 1 & 1    & 1    & 1    \\
\bottomrule
\end{tabular}
\end{minipage}
\end{center}

\section{Constrained EM Implementation Details}
\label{app:constrained-em-details}

\paragraph{Mean constraints.}
After initialization and each emission update, we project
the location of state $k$ onto its box $C_k$ in
\eqref{eq:constrained-box}. The table below gives the bounds on
$\mu_{kj}$, the mean of standardized feature $j$ in state $k$,
with $\varepsilon=0.25$. A blank entry denotes a free coordinate.
Features omitted from the table have free means in every regime.

\begin{center}
\small
\begin{tabular}{@{}lcccc@{}}
\toprule
Feature $j$ & Crisis (0) & Inflation (1) & Expansion (2) & Fragile (3) \\
\midrule
\texttt{SPXT}
  & $\le -2\varepsilon$ & $\le -\varepsilon$
  & $\ge 0$ & $[0,\varepsilon]$ \\
\texttt{VIX}
  & $\ge \varepsilon$ & $\le 0$ & $\le 0$ & $\ge \varepsilon$ \\
\texttt{LUACOAS}
  & $\ge \varepsilon$ & $\le 0$ & $\le 0$ & $\ge \varepsilon$ \\
\texttt{BCOMTR}, \texttt{BCIT1T}
  & & $\ge 2.5\varepsilon$ & & \\
\texttt{DXY}, \texttt{LUATTRUU}
  & & $\le -1.5\varepsilon$ & & \\
\texttt{YC\_10Y\_3M}, \texttt{CU\_AU\_RATIO}
  & & $\ge 1.5\varepsilon$ & & \\
\texttt{LF98TRUU}
  & & & $\ge 0$ & \\
\bottomrule
\end{tabular}
\end{center}

Projection clips each bounded coordinate to its admissible interval.
Mixing weights and HMM transitions follow their
respective model specifications.

\paragraph{Projected emission updates.}
For training month $t$, let $r_{tk}$ be the mixture responsibility
or smoothed HMM probability of state $k$. Set $w_{tk}=r_{tk}u_{tk}$,
where $u_{tk}=1$ for Gaussian emissions and
\eqref{eq:app-mix-u} gives $u_{tk}$ for Student-$t$ emissions.
For a state with positive mass, define
\[
N_k=\sum_t r_{tk},\qquad
W_k=\sum_t w_{tk},\qquad
\widehat\mu_{kj}=\frac{\sum_t w_{tk}x_{tj}}{W_k},\qquad
D_{kj}(m)=\frac{\sum_t w_{tk}(x_{tj}-m)^2}{N_k}.
\]
Here $x_{tj}$ is the standardized observation and $D_{kj}(m)$
is the weighted residual second moment about location $m$. The updated location and diagonal dispersion are
\[
\mu_k^+=\Pi_{C_k}(\widehat\mu_k),\qquad
(\sigma_{kj}^2)^+
=\operatorname{clip}\!\left(D_{kj}(\mu_{kj}^+),L_{kj},U_{kj}\right).
\]
The denominator in $D_{kj}$ is $N_k=\sum_t r_{tk}$ for both
Gaussian and Student-$t$ emissions. For Student-$t$ emissions the
latent scale weight $u_{tk}$ enters the numerator. Before clipping,
the increase in the residual second moment induced by moving the
location to the boundary is
\[
D_{kj}(\mu_{kj}^+)-D_{kj}(\widehat\mu_{kj})
=\frac{W_k}{N_k}(\mu_{kj}^+-\widehat\mu_{kj})^2.
\]\\
The lower scale bound $L_{kj}$ is $10^{-4}$ for mixtures and
$10^{-3}$ for HMMs, except that Crisis has a lower bound of $9$
on \texttt{SPXT}, \texttt{VIX}, and \texttt{LUACOAS}. The upper bound
$U_{kj}$ is $25$ for Crisis and Fragile and $2.25$ for Inflation
and Expansion. Fragile \texttt{SPXT} has the tighter upper bound
$\varepsilon^2=0.0625$.\\
For Gaussian emissions, $\sigma_{kj}^2$ is the conditional variance.
For Student-$t$ emissions with $\nu$ degrees of freedom, it is the
diagonal scale entry and the conditional variance is
$\nu\sigma_{kj}^2/(\nu-2)$.
\paragraph{Implications of the scale bounds.}
The emission bounds help explain what the constraints impose. For
\texttt{SPXT}, \texttt{VIX}, and \texttt{LUACOAS}, the Crisis root scale
$\sigma_{k,j}$ lies between $3$ and $5$, while the Expansion root
scale is at most $1.5$ on every feature. With $\nu=6$, the conditional
standard deviation is $\sqrt{6/(6-2)}\,\sigma_{k,j}$. The corresponding
standard deviations are therefore $3.67$ to $6.12$ for those three
Crisis features and at most $1.84$ for Expansion, in standardized
feature units. Holding other parameters fixed, a larger scale lowers
density at the component location and spreads mass farther into the
tails. These bounds could contribute to the concentration in Expansion.
Assignment in any given month also depends on the fitted locations, the
full feature vector, and the mixture weights or HMM predictive state
probabilities.

\section{OACI-EM Implementation Details}
\label{app:oaci-em-details}

\paragraph{Template-matched initialization and dated anchors.}
At a fresh initialization, a one-to-one assignment matches the initial
centroids to the regime mean templates by maximizing total signed
alignment. The ordered centroids are then projected onto the mean bounds
given below. During training, a calendar anchor becomes eligible once
its date is at least twelve months before the last training date.
Write $k^\star$ for its target state. Soft anchors use $b=0.90$
and hard Crisis anchors use $b=1$.\\
For the GMM and SMM, each anchor acts on a single responsibility vector
$\gamma_t$. If $\gamma_{t,k^\star}\ge b$, then
$\gamma_t^{\mathrm{PR}}=\gamma_t$. Otherwise, posterior regularization
raises the target probability to $b$ and preserves the relative
probabilities of the remaining states:
\[
\gamma_{t,k^\star}^{\mathrm{PR}}=b,\qquad
\gamma_{t,j}^{\mathrm{PR}}
=\gamma_{t,j}\frac{1-b}{1-\gamma_{t,k^\star}}
\quad (j\ne k^\star).
\]\\
For the GHMM and SHMM, anchors modify the log emission potentials
$\ell_{t,k}$ before forward-backward inference. A hard Crisis anchor
retains the target log density and sets every other potential on the
log scale to $-10^{300}$. At a soft anchor $a$, dated $t_a$
with target $k_a^\star$, the adjusted target potential is
\[
\ell'_{t_a,k_a^\star}=\ell_{t_a,k_a^\star}+\lambda_a,
\qquad 0\le\lambda_a\le80.
\]
Each restart initializes the multipliers at zero and carries them
forward between EM iterations within that restart. Within an E-step, a
single forward-backward pass at the current potentials gives all
shortfalls $d_a=b-\gamma_{t_a,k_a^\star}$. The solver updates the anchors
with $d_a>10^{-3}$ simultaneously. On the first inner iteration, if all
multipliers are zero, it sets each of those multipliers to
$\log(b/(1-b))=\log 9$. Otherwise it adds $\max(2.5d_a,0.75)$, capping
each multiplier at 80. The other multipliers keep their current values.
The solver repeats this step until the largest shortfall is at most
$10^{-3}$ or it reaches twelve iterations. A soft anchor can therefore
finish below $0.90$. The adjustment can affect neighboring months
through the joint state-path distribution.\\
A final forward-backward pass gives the adjusted path distribution
$q^{\mathrm{PR}}$, with state marginals
$\gamma_{t,j}^{\mathrm{PR}}=q^{\mathrm{PR}}(z_t=j)$ and transition
marginals $\xi_t^{\mathrm{PR}}(j,k)
=q^{\mathrm{PR}}(z_t=j,z_{t+1}=k)$. They satisfy
\[
\sum_k\xi_t^{\mathrm{PR}}(j,k)=\gamma_{t,j}^{\mathrm{PR}},
\qquad
\sum_j\xi_t^{\mathrm{PR}}(j,k)=\gamma_{t+1,k}^{\mathrm{PR}}.
\]
The emission update uses $\gamma_t^{\mathrm{PR}}$, and the initial-state
update uses $\gamma_1^{\mathrm{PR}}$. The transition update uses
$N_{jk}=\sum_{t=1}^{T-1}\xi_t^{\mathrm{PR}}(j,k)$ with the same
Dirichlet prior as the other HMM specifications
(Appendix~\ref{app:baseline-models}).

\paragraph{Location and scale updates.}
For mixtures and HMMs, let $r_{tk}$ be the anchor-adjusted training
probability of state $k$ and $w_{tk}=r_{tk}u_{tk}$, as in
Appendix~\ref{app:constrained-em-details}. Here $u_{tk}=1$ for Gaussian
emissions and is the conditional scale weight in \eqref{eq:app-mix-u}
for Student-$t$ emissions. Write $N_k=\sum_t r_{tk}$ and
$W_k=\sum_t w_{tk}$. For $W_k>0$, the weighted mean and its
template adjustment are
\[
\widehat\mu_{kj}=\frac{\sum_t w_{tk}x_{tj}}{W_k},
\qquad
\bar\mu_{kj}=
\frac{W_k\widehat\mu_{kj}+\kappa_{kj}t_{kj}}{W_k+\kappa_{kj}}.
\]
On an active coordinate, the mean template sign $s^\mu_{kj}$ is
nonzero and the target is $t_{kj}=\varepsilon s^\mu_{kj}$, with
$\varepsilon=0.25$. The Expansion target for \texttt{SPXT} is
$0.4\varepsilon$. The default strength is $\kappa_{kj}=\kappa=100$.
It doubles for \texttt{SPXT}, \texttt{VIX}, and \texttt{LUACOAS},
with two exceptions: Expansion \texttt{SPXT} uses $\kappa$, and
Fragile \texttt{VIX} and \texttt{LUACOAS} use $4\kappa$.
Inactive coordinates retain $\widehat\mu_{kj}$, except for Inflation
\texttt{VIX} and \texttt{LUACOAS}, which shrink toward zero:
\[
\bar\mu_{kj}=\frac{W_k\widehat\mu_{kj}}{W_k+4\kappa}.
\]\\
Mixtures and HMMs estimate scales about $\bar\mu_{kj}$, then project
the means without recomputing the scales. The projection uses only these six bounds:
\begin{align*}
\mu_{\mathrm{Crisis},\texttt{SPXT}} &\le -2\varepsilon,
& \mu_{\mathrm{Crisis},\texttt{VIX}} &\ge \varepsilon,\\
\mu_{\mathrm{Inflation},\texttt{VIX}} &\le 0,
& \mu_{\mathrm{Inflation},\texttt{LUACOAS}} &\le 0,\\
\mu_{\mathrm{Inflation},\texttt{BCOMTR}} &\ge 2.5\varepsilon,
& \mu_{\mathrm{Inflation},\texttt{BCIT1T}} &\ge 2.5\varepsilon.
\end{align*}\\
The mixture M-step caps Fragile at $0.18$, redistributes excess mass,
then caps Inflation at $0.22$ and redistributes again. The second
redistribution can raise Fragile above $0.18$. Among the 37 January
refits from 1990 through 2026, the final Fragile weight exceeded
$0.18$ once in the GMM and three times in the SMM, reaching $0.1918$
and $0.1880$, respectively. Inflation respected its cap in all
audited mixture fits.\\
For mixtures and HMMs, the diagonal scale on feature $j$ in state
$k$ is the product of a state scale $\tau_k$ and a shared feature
shape $v_j$ before flooring. Given the weights and locations, the
residual moment and the scale contribution to the negative expected
complete-data log likelihood, up to a constant, are
\[
D_{kj}=\frac{1}{N_k}\sum_t w_{tk}(x_{tj}-\bar\mu_{kj})^2,
\qquad
\mathcal L(\tau,v)=\frac12\sum_{k,j}N_k
\left\{\log(\tau_kv_j)+\frac{D_{kj}}{\tau_kv_j}\right\}.
\]
Set $\delta=10^{-4}$ for mixtures and $10^{-3}$ for HMMs.
Initialize $v_j=\max\{\delta,\sum_k N_kD_{kj}/\sum_k N_k\}$.
Each of twelve sweeps updates the state scales, enforces the Crisis
ordering, and updates the feature shape:
\begin{align*}
a_k &= \max\left\{10^{-8},\,\frac1d\sum_{j=1}^d
                       \frac{D_{kj}}{v_j}\right\},\\
\tau &= \operatorname*{arg\,min}_{b:\,b_0\ge b_k\ \forall k}
                    \sum_k N_k(b_k-a_k)^2,\\
v_j &= \max\left\{\delta,\,
            \frac{\sum_k N_kD_{kj}/\tau_k}{\sum_k N_k}\right\}.
\end{align*}
Here $d$ is the number of features. The projection pools Crisis
($k=0$) with the largest violating scales at their mass-weighted
mean. In the absence of numerical floors, the ratio updates minimize
$\mathcal L$ over each factor with the other held fixed, subject to
the Crisis ordering in the state-scale update.\\
The stored entries are
$(\sigma_{kj}^2)^+=\max\{\delta,\tau_kv_j\}$.
Only the products identify the fitted scales. For reporting,
$g=(\prod_jv_j)^{1/d}$, $v_j^\circ=v_j/g$, and
$\tau_k^\circ=g\tau_k$ give $\prod_jv_j^\circ=1$ while preserving
every product. Estimation uses the unnormalized factors. An active
entrywise floor can break exact proportionality, and the conditional
scale updates alone give no likelihood-ascent guarantee for OACI-EM
as a whole.

\paragraph{Out-of-sample scoring.}
Let $S=(s^\mu_{kj})\in\{-1,0,+1\}^{K\times d}$ be the mean sign
matrix and $J_k=\{j:s^\mu_{kj}\ne0\}$ the active features for
state $k$. For the standardized observation $x_t$, the alignment
in \eqref{eq:oaci-tilt} is
\[
\mathrm{align}(x_t,s_k^\mu)
=\frac{1}{|J_k|}\sum_{j\in J_k}s^\mu_{kj}x_{t,j}.
\]
With $\beta=0.4$, the tilt reweights mixture responsibilities and
filtered HMM probabilities. No calendar anchor applies to the month
being scored.

\section{Walk-Forward and Smoothing Details}
\label{app:walk-forward-details}

\paragraph{Expanding-window loop (mixtures and HMMs).}

For each month $t$ from January 1990 through the end of the sample, the engine (i)~forms $\mathcal{D}_t=\{x_s:s<t\}$, (ii)~standardizes training rows and month $t$ using causal EWM moments, (iii)~fits the selected model-labeling pair on the training sample, and (iv)~scores month $t$. Mixtures compute component responsibilities conditional on $x_t$. HMMs first form the one-step predictive distribution $\Pr(z_t=k\mid x_{1:t-1})$, then incorporate $x_t$ to obtain the filtered posterior $\Pr(z_t=k\mid x_{1:t})$. OACI-EM applies the template tilt \eqref{eq:oaci-tilt} to the resulting probabilities. The adjusted scores are subsequently blended according to \eqref{eq:signal-blend}. Windows with fewer than 40 training months are skipped.

\paragraph{Blending regime scores and constructing the discrete signal.}
All mixture and HMM specifications use the same post-processing rules:
\begin{itemize}
  \item EMA smooth \eqref{eq:signal-blend} with span $S=8$ and weight on the current score $w=0.35$;
  \item minimum dwell $d=4$ months and switch threshold $\delta=0.40$ on the smoothed-regime scores;
  \item Crisis fast entry when $g_t(\mathrm{Crisis})\ge 0.80$;
  \item Crisis fast exit when $g_t(\mathrm{Crisis})<0.20$.
\end{itemize}
The dwell and Crisis rules determine the discrete trading signal. Score-weighted allocations use $\tilde g_t$ directly.

\section{Hyperparameters}

The Student-$t$ models use fixed degrees of freedom $\nu=6$. For comparability, the reported mixtures and HMMs use a common score blend and discrete switching rule, specified in Appendix~\ref{app:walk-forward-details}.\\
The economic templates, OACI settings, and historical anchor calendar are
research design choices. The calendar was constructed retrospectively.
Within each walk-forward fit, an anchor becomes eligible only twelve months
after its reference month.

\section{Sign Templates}
\begin{table}[H]
\centering
\caption{Sign templates by feature and regime (mean / volatility).
Entries are $+1$, $0$, or $-1$: the expected direction of the regime mean
(resp.\ relative volatility) in standardized feature space; $0$ leaves the
coordinate unrestricted.}
\label{tab:sign-templates}

\begin{tabular}{@{}lcccc@{}}
\toprule
Feature & Crisis & Inflation & Expansion & Fragile \\
\midrule
SPXT         & $-1$ / $+1$ & $-1$ / $0$  & $+1$ / $-1$ & $+1$ / $+1$ \\
VIX          & $+1$ / $+1$ & $0$  / $0$  & $-1$ / $-1$ & $+1$ / $+1$ \\
LUACOAS      & $+1$ / $+1$ & $0$  / $0$  & $-1$ / $-1$ & $+1$ / $+1$ \\
MXEF         & $-1$ / $+1$ & $0$  / $0$  & $0$  / $-1$ & $+1$ / $+1$ \\
BCOMTR       & $-1$ / $+1$ & $+1$ / $+1$ & $-1$ / $-1$ & $0$  / $0$ \\
LUATTRUU     & $+1$ / $+1$ & $-1$ / $+1$ & $+1$ / $-1$ & $0$  / $+1$ \\
LF98TRUU     & $-1$ / $+1$ & $0$  / $0$  & $+1$ / $-1$ & $0$  / $+1$ \\
DXY          & $+1$ / $+1$ & $-1$ / $+1$ & $0$  / $0$  & $0$  / $+1$ \\
BCIT1T       & $-1$ / $+1$ & $+1$ / $+1$ & $0$  / $-1$ & $0$  / $+1$ \\
NEIXCTAT     & $+1$ / $+1$ & $0$  / $0$  & $0$  / $0$  & $-1$ / $+1$ \\
CU\_AU\_RATIO & $-1$ / $+1$ & $+1$ / $+1$ & $+1$ / $-1$ & $0$  / $+1$ \\
YC\_10Y\_3M  & $-1$ / $+1$ & $+1$ / $+1$ & $0$  / $0$  & $-1$ / $+1$ \\
\bottomrule
\end{tabular}
\end{table}

\begin{table}[H]
\centering
\caption{Fourteen dated training exemplars used by OACI. Soft anchors
impose a target-state responsibility floor in mixtures and target a
marginal probability of 0.90 through a capped iteration in HMMs. Bold denotes a hard
Crisis anchor. An exemplar becomes eligible at least twelve months
after its date and receives no calendar supervision when first scored.}
\label{tab:anchor-calendar}
\begin{tabular}{@{}llp{7.2cm}@{}}
\toprule
Date & Regime & Justification \\
\midrule
Sep 2008 & Crisis     & Lehman Brothers collapse \\
Oct 2008 & \textbf{Crisis} & GFC, acute aftermath of Lehman \\
Nov 2008 & \textbf{Crisis} & GFC, continued market freeze \\
Mar 2020 & \textbf{Crisis} & COVID-19 crash \\
Sep 1974 & Inflation  & Oil-crisis stagflation \\
Mar 2022 & Inflation  & Start of the Fed hiking cycle \\
Jun 2022 & Inflation  & US CPI near its cycle peak ($\approx 9\%$) \\
Sep 2022 & Inflation  & Continued rate-shock inflation \\
Jun 1997 & Expansion  & Late-1990s calm bull market \\
Jun 2005 & Expansion  & Mid-2000s low-volatility expansion \\
Jun 2016 & Expansion  & Post-GFC low-volatility bull market \\
Aug 2011 & Fragile    & US credit downgrade / European sovereign stress \\
Aug 2015 & Fragile    & China devaluation / EM growth scare \\
Dec 2018 & Fragile    & Q4 2018 growth scare and equity sell-off \\
\bottomrule
\end{tabular}
\end{table}

\section{Additional Walk-Forward Timelines}
\label{app:additional-timelines}

\begin{figure}[H]
\centering
\articlegraphic[width=\textwidth]{ 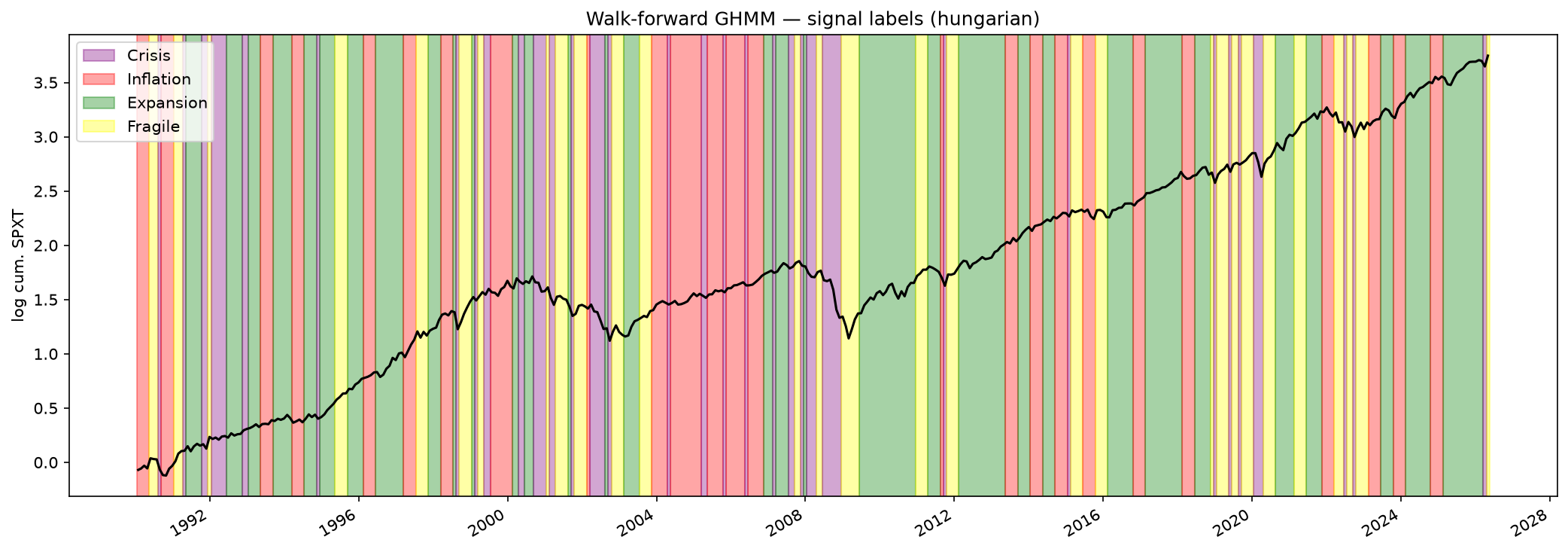}\\[0.6em]
\articlegraphic[width=\textwidth]{ 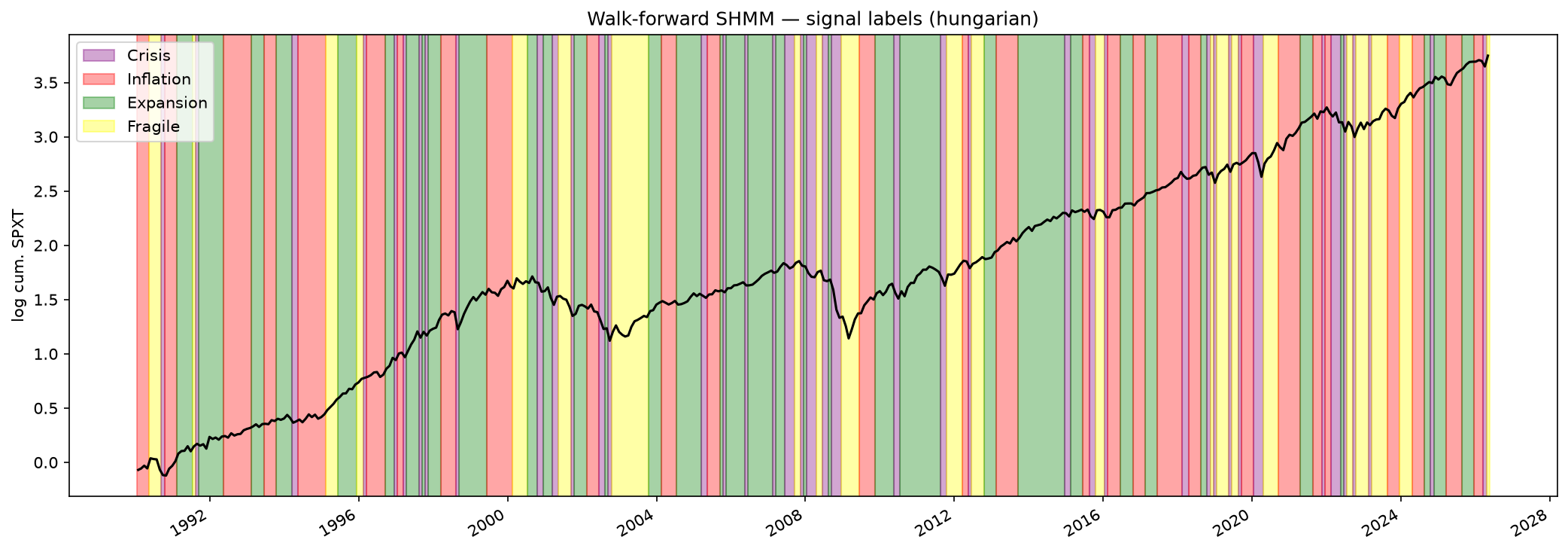}
\caption{Walk-forward signal labels under Hungarian matching. Top: GHMM.
Bottom: SHMM.}
\label{fig:app-hungarian-hmms}
\end{figure}

\paragraph{Hungarian matching on HMMs.}
Over 435 monthly transitions, the Hungarian discrete trading signal changes state 118 times for the GHMM and 126 times for the SHMM (Figure~\ref{fig:app-hungarian-hmms}).

\begin{figure}[H]
\centering
\articlegraphic[width=\textwidth]{ 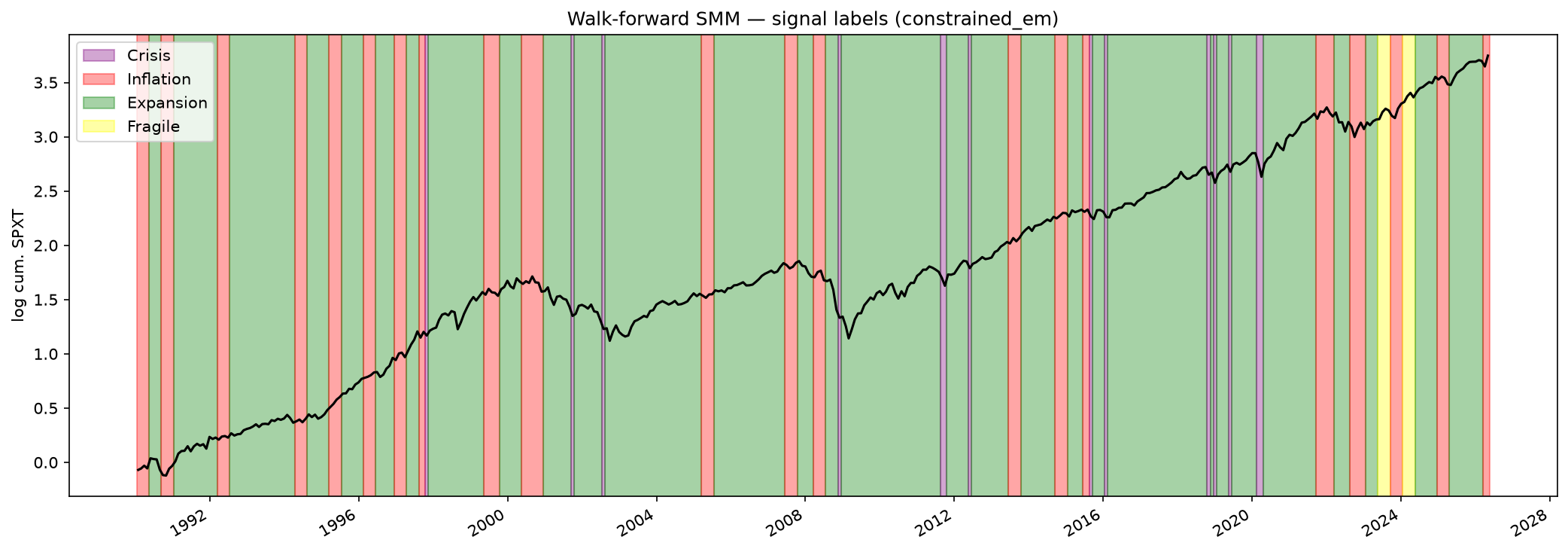}
\caption{Walk-forward signal labels under Constrained EM for the SMM.}
\label{fig:app-constrained-smm}
\end{figure}

\begin{figure}[H]
\centering
\articlegraphic[width=\textwidth]{ 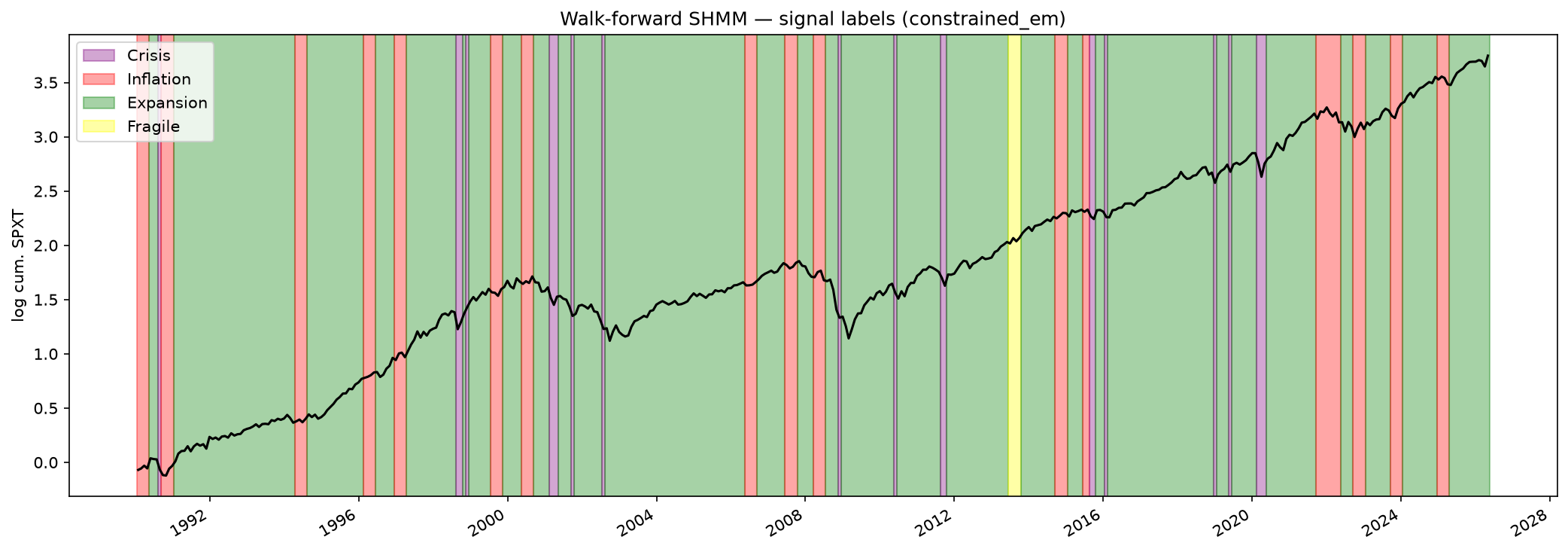}
\caption{Walk-forward signal labels under Constrained EM for the SHMM.}
\label{fig:app-constrained-shmm}
\end{figure}

\paragraph{Constrained EM on Student-$t$ models.}
Expansion accounts for 75.5\% of SMM months and 79.1\% of SHMM months under Constrained EM. Fragile accounts for 1.8\% and 0.9\%, respectively (Figures~\ref{fig:app-constrained-smm} and~\ref{fig:app-constrained-shmm}).

\begin{figure}[H]
\centering
\articlegraphic[width=\textwidth]{ 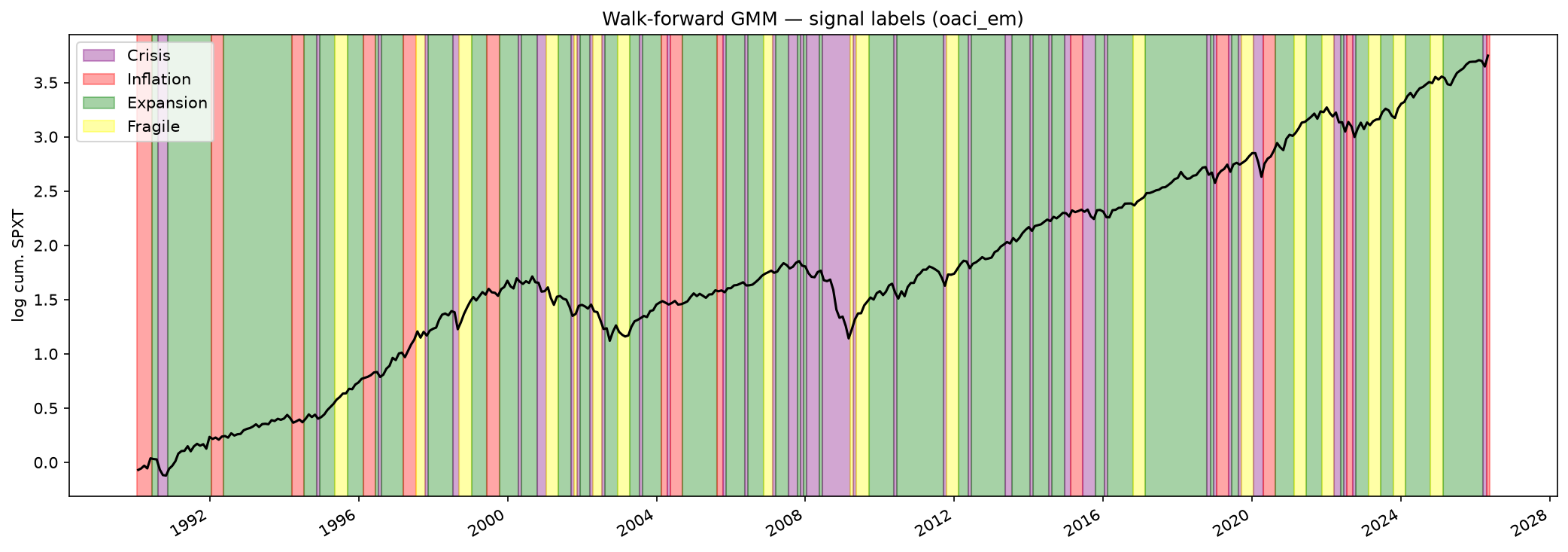}\\[0.6em]
\articlegraphic[width=\textwidth]{ 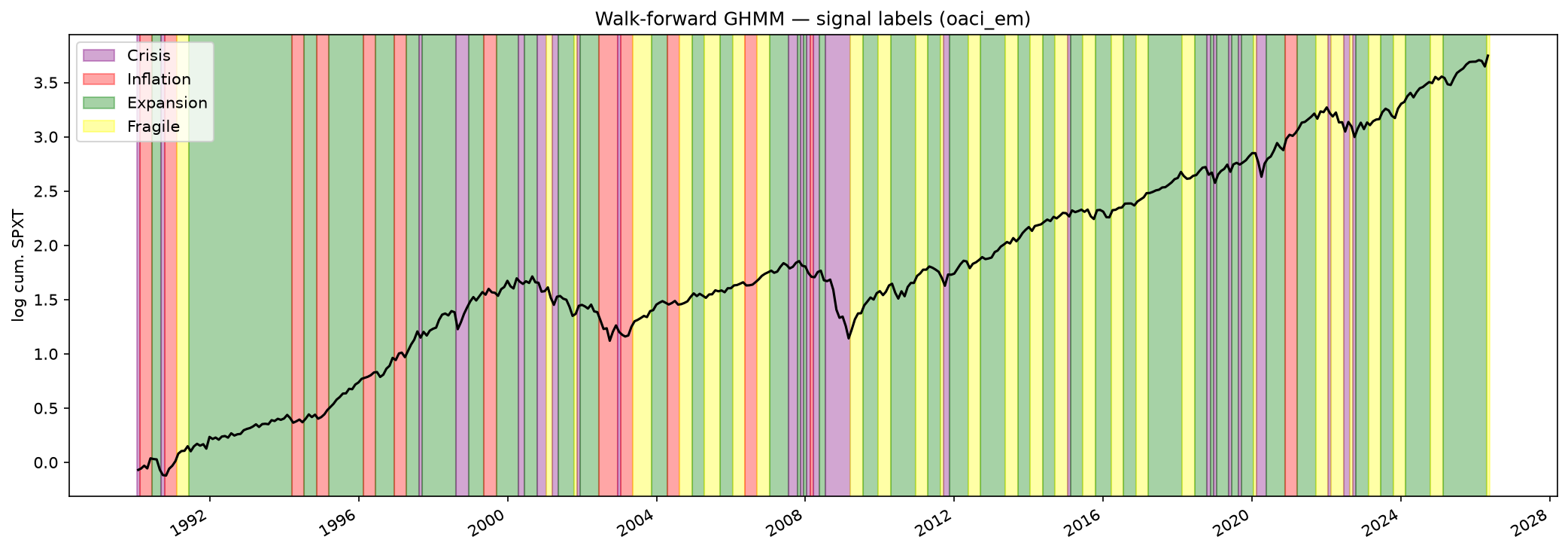}
\caption{Walk-forward signal labels under OACI-EM. Top: GMM. Bottom: GHMM.}
\label{fig:app-oaci-gaussian}
\end{figure}

\paragraph{OACI-EM on Gaussian models.}
All four labels appear under OACI in both Gaussian models. Expansion accounts for 59.6\% of GMM months and 55.5\% of GHMM months (Figure~\ref{fig:app-oaci-gaussian}).

\section{Backtest Metrics}
\label{app:backtest-metrics}

Let $a_t=w_{t-1}$ denote the executed target and $d_t$ the risky-asset weights after market drift, immediately before trading. The portfolio starts in cash. At a cost of $c=0.0005$ per dollar bought or sold, the fraction $s_t$ of pre-trade wealth remaining after costs satisfies
\[
s_t=1-cT_t.
\]
Turnover per dollar of pre-trade wealth is
\begin{equation}
T_t=\sum_j |s_t a_{tj}-d_{tj}|.
\end{equation}
With zero interest on cash, the net return is $s_t(1+a_t^\top r_t^{\rm asset})-1$, where $r_t^{\rm asset}$ is the vector of monthly risky-asset returns. Strategies and benchmarks pay the same rate on the initial purchase and subsequent rebalancing trades.\\
For net monthly returns $r_t$, CAGR is $\bigl[\prod_{t=1}^{n}(1+r_t)\bigr]^{12/n}-1$, annualized volatility is $\sqrt{12}\,s(r)$, and the annualized Sharpe ratio ($r_f=0$) is $\sqrt{12}\,\bar r/s(r)$. The US dollar exposure uses DXY index returns and excludes funding, carry, and roll costs. Starting from $V_0=1$, wealth is $V_t=\prod_{s=1}^t(1+r_s)$. The drawdown path is $D_t=V_t/\max_{0\leq u\leq t}V_u-1$. The maximum drawdown reported as a loss magnitude is $-\min_t D_t$.

\section{Allocations using the discrete trading signal}
\label{app:hard-signals}

Under single-regime SMM all-weather allocation, Hungarian matching records 0.95 and OACI records 0.76 (Table~\ref{tab:app-hard-main}).

\begin{table}[H]
\centering
\caption{Annualized Sharpe ratios ($r_f=0$) under single-regime allocation for the bond-floor and all-weather rules.}
\label{tab:app-hard-main}
\begin{tabular}{@{}lccc@{\hspace{1.5em}}ccc@{}}
\toprule
& \multicolumn{3}{c}{Bond floor} & \multicolumn{3}{c}{All-weather} \\
\cmidrule(lr){2-4}\cmidrule(lr){5-7}
Model & Hun. & Constr. & OACI & Hun. & Constr. & OACI \\
\midrule
GMM & 0.93 & 0.85 & 0.86 & 0.72 & 0.67 & 0.76 \\
SMM & 0.93 & 0.85 & 0.96 & 0.95 & 0.63 & 0.76 \\
GHMM & 0.99 & 0.95 & 0.74 & 0.71 & 0.90 & 0.59 \\
SHMM & 0.93 & 0.87 & 1.07 & 0.73 & 0.65 & 0.81 \\
\bottomrule
\end{tabular}
\end{table}

\section{Additional rules-based strategies}
\label{app:handcrafted-strategies}

These rules use the discrete and score-weighted allocations and transaction costs defined in Section~\ref{sec:handcrafted}.

\paragraph{Risk-on / risk-off.}
The discrete signal holds equities in Expansion and cash otherwise. When using the smoothed regime scores, the equity weight is $\tilde g_t(\mathrm{Expansion})$.

\paragraph{Safe haven.}
The equity weight follows the same rule, and Treasuries absorb the balance. When using the smoothed regime scores, $w_t^{\mathrm{eq}}=\tilde g_t(\mathrm{Expansion})$ and $w_t^{\mathrm{bonds}}=1-w_t^{\mathrm{eq}}$.

\paragraph{Cash shelter.}
Let $p_{E,t}=\tilde g_t(\mathrm{Expansion})$ and
$p_{R,t}=\tilde g_t(\mathrm{Crisis})+\tilde g_t(\mathrm{Fragile})$. The
blended allocation sets
\[
\begin{aligned}
w_t^{\mathrm{eq}} &= 0.7p_{E,t},\\
w_t^{\mathrm{cash}} &= (1-w_t^{\mathrm{eq}})\,p_{R,t}(0.8+0.2p_{R,t}),\\
w_t^{\mathrm{bonds}} &= 1-w_t^{\mathrm{eq}}-w_t^{\mathrm{cash}}.
\end{aligned}
\]
With the discrete signal, the rule holds $70\%$ equities and $30\%$
Treasuries in Expansion, all Treasuries in Inflation, and all cash in
Crisis or Fragile.\\
Across the twelve specifications using blended weights, GMM--OACI has the highest sample Sharpe ratio for risk-on/off (0.91) and cash shelter (1.18). For safe haven, SHMM--OACI ranks highest (1.17). Table~\ref{tab:app-remaining-soft} reports all three comparisons.

\begin{table}[H]
\centering
\caption{Annualized Sharpe ratios ($r_f=0$) for the remaining rules-based strategies using score-weighted allocations.}
\label{tab:app-remaining-soft}
\begin{tabular}{@{}lccc@{\hspace{1.2em}}ccc@{\hspace{1.2em}}ccc@{}}
\toprule
& \multicolumn{3}{c}{Risk-on/off} & \multicolumn{3}{c}{Safe haven} & \multicolumn{3}{c}{Cash shelter} \\
\cmidrule(lr){2-4}\cmidrule(lr){5-7}\cmidrule(lr){8-10}
Model & Hun. & Constr. & OACI & Hun. & Constr. & OACI & Hun. & Constr. & OACI \\
\midrule
GMM & 0.78 & 0.82 & 0.91 & 1.08 & 1.05 & 1.12 & 1.11 & 1.09 & 1.18 \\
SMM & 0.66 & 0.82 & 0.82 & 1.02 & 0.96 & 1.16 & 1.01 & 1.08 & 1.14 \\
GHMM & 0.80 & 0.82 & 0.86 & 1.10 & 0.98 & 1.07 & 1.11 & 1.06 & 1.15 \\
SHMM & 0.62 & 0.78 & 0.87 & 1.01 & 0.90 & 1.17 & 1.05 & 1.02 & 1.17 \\
\bottomrule
\end{tabular}
\end{table}
\subsection{Bond-floor path}
\begin{figure}[H]
\centering
\articlegraphic[width=\textwidth]{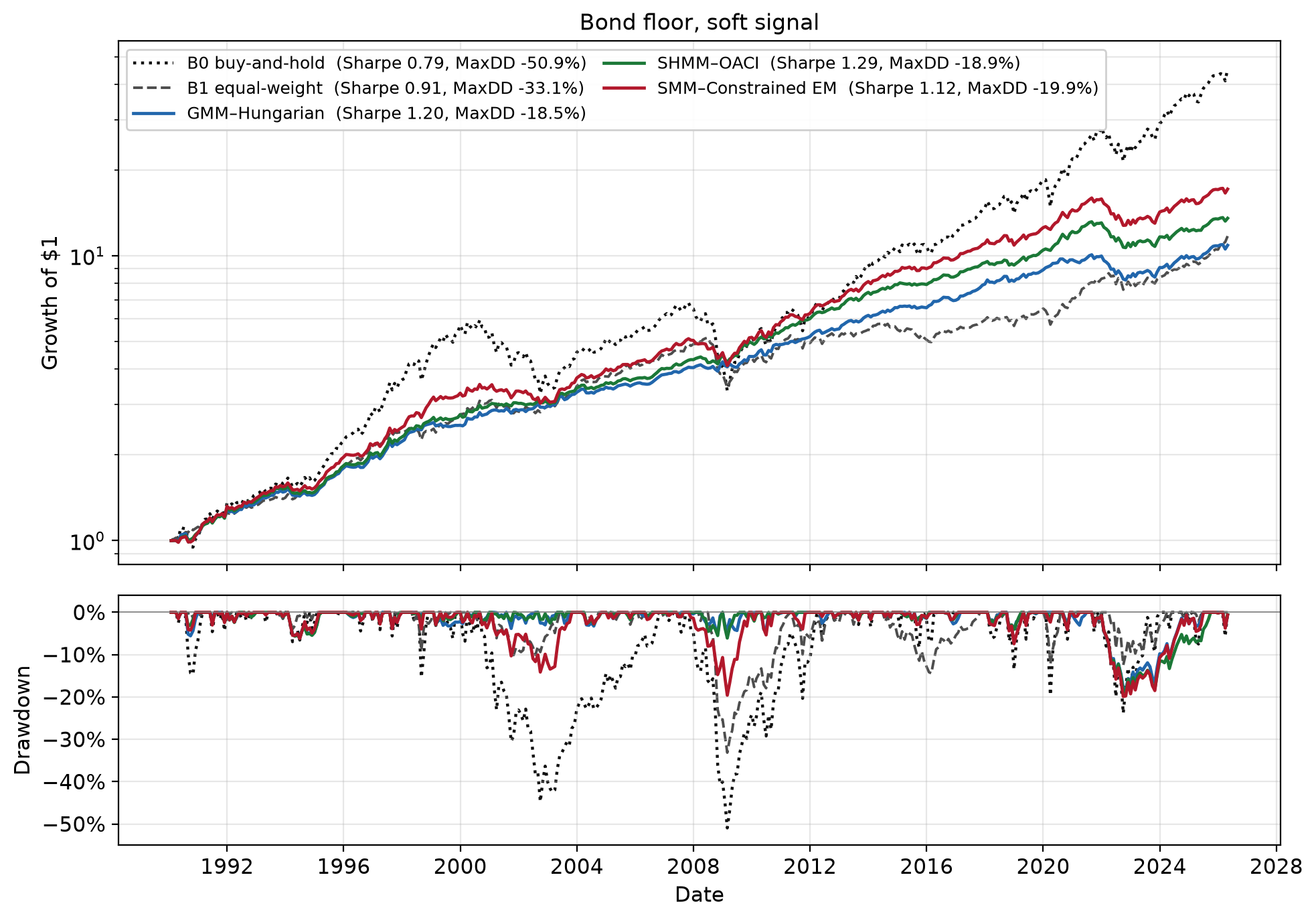}
\caption{Low-risk bond floor under blended allocation. The panels show
growth of \$1 on a log scale and drawdown for B0, B1, GMM-Hungarian,
SHMM-OACI, and SMM-Constrained EM.}
\label{fig:bond-floor-path}
\end{figure}

\section{Fixed regime portfolios for the three- and seven-asset universes}
\label{app:data-driven-sleeves}

All regime portfolios in this appendix are estimated once from a static GMM--OACI fit over March 1971-December 1989. At each growth setting ($\eta=0,20,40$), all specifications use the same four portfolios throughout January 1990-April 2026. Their blended regime scores $\tilde g_t$ determine the monthly allocation.

\paragraph{Three-asset regime portfolios.}
Inflation holds commodities at every growth setting. Expansion holds only equities at $\eta=20$ and $40$, and Fragile reaches full equity exposure at $\eta=40$ (Table~\ref{tab:dd-sleeves-3}).

\begin{table}[H]
\centering
\footnotesize
\setlength{\tabcolsep}{1.5pt}
\caption{Three-asset regime portfolios by growth setting ($\eta \in \{0, 20, 40\}$).}
\label{tab:dd-sleeves-3}
\begin{tabular}{@{}lr@{\hspace{1em}}rrr@{\hspace{1em}}rrr@{\hspace{1em}}rrr@{}}
\toprule
& & \multicolumn{3}{c}{Low ($\eta=0$)} & \multicolumn{3}{c}{Medium ($\eta=20$)} & \multicolumn{3}{c}{High ($\eta=40$)} \\
\cmidrule(lr){3-5}\cmidrule(lr){6-8}\cmidrule(lr){9-11}
Regime & $n$ & SPXT & LUATTRUU & BCOMTR & SPXT & LUATTRUU & BCOMTR & SPXT & LUATTRUU & BCOMTR \\
\midrule
Crisis & 17 & 0.06 & 0.94 & 0.00 & 0.00 & 1.00 & 0.00 & 0.00 & 1.00 & 0.00 \\
Inflation & 28 & 0.00 & 0.00 & 1.00 & 0.00 & 0.00 & 1.00 & 0.00 & 0.00 & 1.00 \\
Expansion & 127 & 0.12 & 0.74 & 0.13 & 1.00 & 0.00 & 0.00 & 1.00 & 0.00 & 0.00 \\
Fragile & 18 & 0.29 & 0.65 & 0.05 & 0.35 & 0.65 & 0.00 & 1.00 & 0.00 & 0.00 \\
\bottomrule
\end{tabular}
\end{table}

\paragraph{Performance of the three-asset portfolios.}
Table~\ref{tab:dd-sharpe-3} reports the three-asset results at all three growth settings. At low growth emphasis ($\eta=0$), the Student-$t$ mixture with OACI-EM achieved the highest sample Sharpe ratio in both asset universes, at 1.20 for three assets and 1.32 for seven. The corresponding equal-weight benchmarks had Sharpe ratios of 0.91 and 0.93.
\begin{table}[H]
\centering
\caption{Annualized Sharpe ratios ($r_f=0$) for three-asset portfolios using score-weighted allocations. The equal-weight three-asset benchmark (B1) records 0.91.}
\label{tab:dd-sharpe-3}
\begin{tabular}{@{}lccc@{\hspace{1.2em}}ccc@{\hspace{1.2em}}ccc@{}}
\toprule
& \multicolumn{3}{c}{Low ($\eta=0$)}
& \multicolumn{3}{c}{Medium ($\eta=20$)}
& \multicolumn{3}{c}{High ($\eta=40$)} \\
\cmidrule(lr){2-4}\cmidrule(lr){5-7}\cmidrule(lr){8-10}
Model & Hun. & Constr. & OACI & Hun. & Constr. & OACI & Hun. & Constr. & OACI \\
\midrule
GMM & 1.00 & 1.09 & 1.16 & 1.02 & 1.01 & 1.12 & 1.00 & 1.00 & 1.08 \\
SMM & 1.09 & 1.04 & 1.20 & 1.07 & 0.92 & 1.18 & 1.02 & 0.91 & 1.12 \\
GHMM & 0.95 & 1.05 & 1.10 & 1.02 & 0.94 & 1.05 & 1.00 & 0.95 & 1.01 \\
SHMM & 1.00 & 0.98 & 1.15 & 0.99 & 0.84 & 1.17 & 0.97 & 0.85 & 1.12 \\
\bottomrule
\end{tabular}
\end{table}

\paragraph{Seven-asset sleeves.}
Tables~\ref{tab:dd-sleeves-8-low}-\ref{tab:dd-sleeves-8-high}
report the portfolio weights used in Section~\ref{sec:data-driven}.
The counts $n$ are the training observations assigned to each regime
in the portfolio estimation. The investable columns are identical across tiers.
Weights are rounded to two decimal places and may sum to $0.99$
or $1.01$.

\begin{table}[H]
\centering
\caption{Seven-asset regime portfolio weights at $\eta=0$.}
\label{tab:dd-sleeves-8-low}
\resizebox{\textwidth}{!}{\begin{tabular}{@{}lrrrrrrrr@{}}
\toprule
Regime & $n$ & SPXT & LUATTRUU & LF98TRUU & BCOMTR
       & DXY & MXEF & NEIXCTAT \\
\midrule
Crisis & 17 & 0.00 & 0.55 & 0.00 & 0.04 & 0.37 & 0.04 & 0.00 \\
Inflation & 28 & 0.00 & 0.00 & 0.00 & 1.00 & 0.00 & 0.00 & 0.00 \\
Expansion & 127 & 0.01 & 0.46 & 0.20 & 0.10 & 0.09 & 0.13 & 0.00 \\
Fragile & 18 & 0.13 & 0.50 & 0.00 & 0.08 & 0.04 & 0.21 & 0.05 \\
\bottomrule
\end{tabular}}
\end{table}

\begin{table}[H]
\centering
\caption{Seven-asset regime portfolio weights at $\eta=20$.}
\resizebox{\textwidth}{!}{\begin{tabular}{@{}lrrrrrrrr@{}}
\toprule
Regime & $n$ & SPXT & LUATTRUU & LF98TRUU & BCOMTR
       & DXY & MXEF & NEIXCTAT \\
\midrule
Crisis & 17 & 0.00 & 0.51 & 0.00 & 0.00 & 0.49 & 0.00 & 0.00 \\
Inflation & 28 & 0.00 & 0.00 & 0.00 & 1.00 & 0.00 & 0.00 & 0.00 \\
Expansion & 127 & 0.32 & 0.00 & 0.00 & 0.00 & 0.00 & 0.68 & 0.00 \\
Fragile & 18 & 0.15 & 0.54 & 0.00 & 0.06 & 0.00 & 0.23 & 0.02 \\
\bottomrule
\end{tabular}}
\end{table}

\begin{table}[H]
\centering
\caption{Seven-asset regime portfolio weights at $\eta=40$.}
\label{tab:dd-sleeves-8-high}
\resizebox{\textwidth}{!}{\begin{tabular}{@{}lrrrrrrrr@{}}
\toprule
Regime & $n$ & SPXT & LUATTRUU & LF98TRUU & BCOMTR
       & DXY & MXEF & NEIXCTAT \\
\midrule
Crisis & 17 & 0.00 & 0.00 & 0.00 & 0.00 & 1.00 & 0.00 & 0.00 \\
Inflation & 28 & 0.00 & 0.00 & 0.00 & 1.00 & 0.00 & 0.00 & 0.00 \\
Expansion & 127 & 0.21 & 0.00 & 0.00 & 0.00 & 0.00 & 0.79 & 0.00 \\
Fragile & 18 & 0.43 & 0.00 & 0.00 & 0.00 & 0.00 & 0.57 & 0.00 \\
\bottomrule
\end{tabular}}
\end{table}
\subsection{Portfolio paths at medium and high growth emphasis}
\begin{figure}[H]
\centering
\articlegraphic[width=\textwidth]{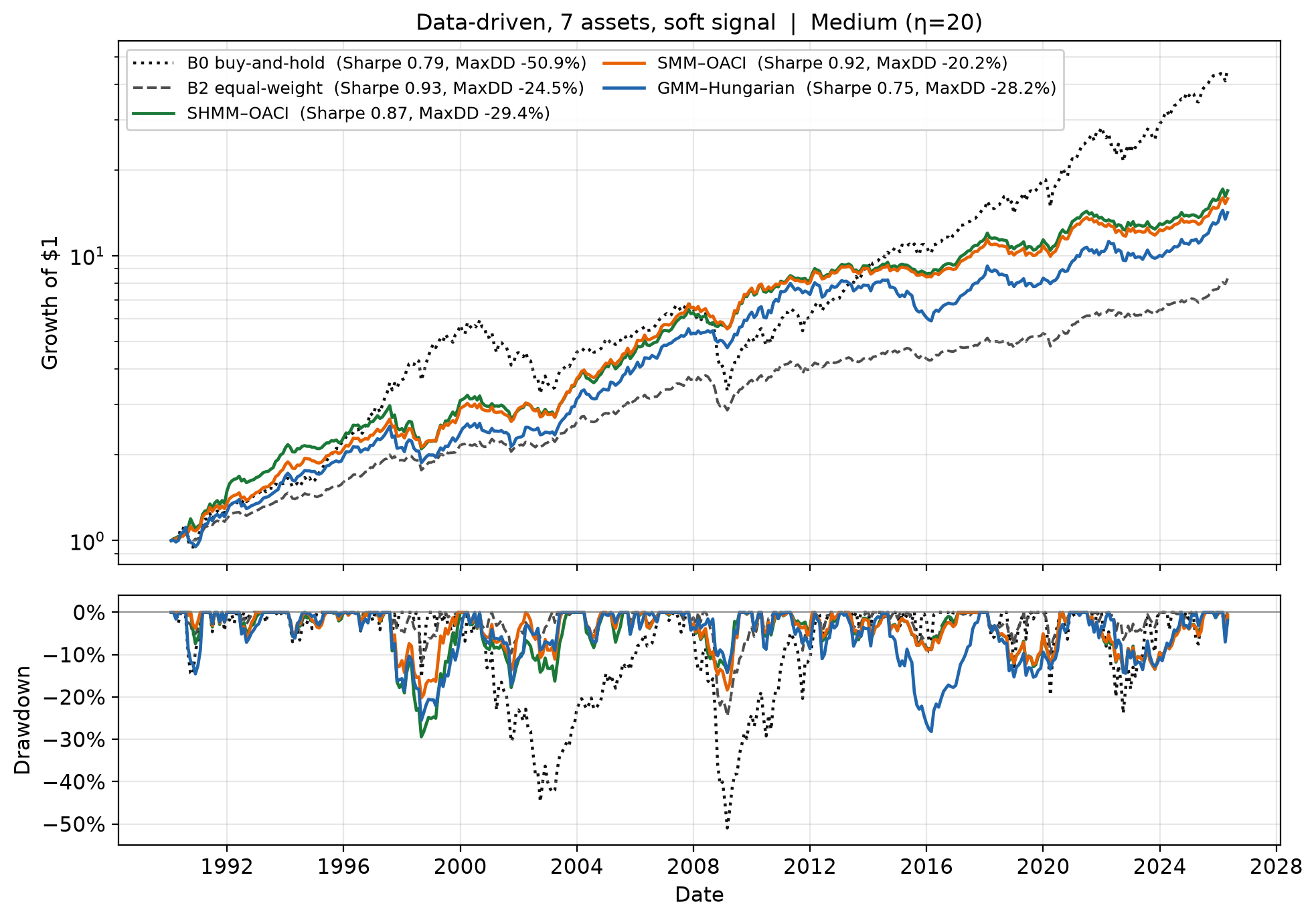}
\caption{Seven-asset portfolios with medium growth emphasis ($\eta=20$). Growth of
\$1 on a log scale and drawdown for the strategies and benchmarks in
Figure~\ref{fig:datadriven-eight-path-low}.}
\label{fig:datadriven-eight-path-medium}
\end{figure}

\begin{figure}[H]
\centering
\articlegraphic[width=0.9\textwidth]{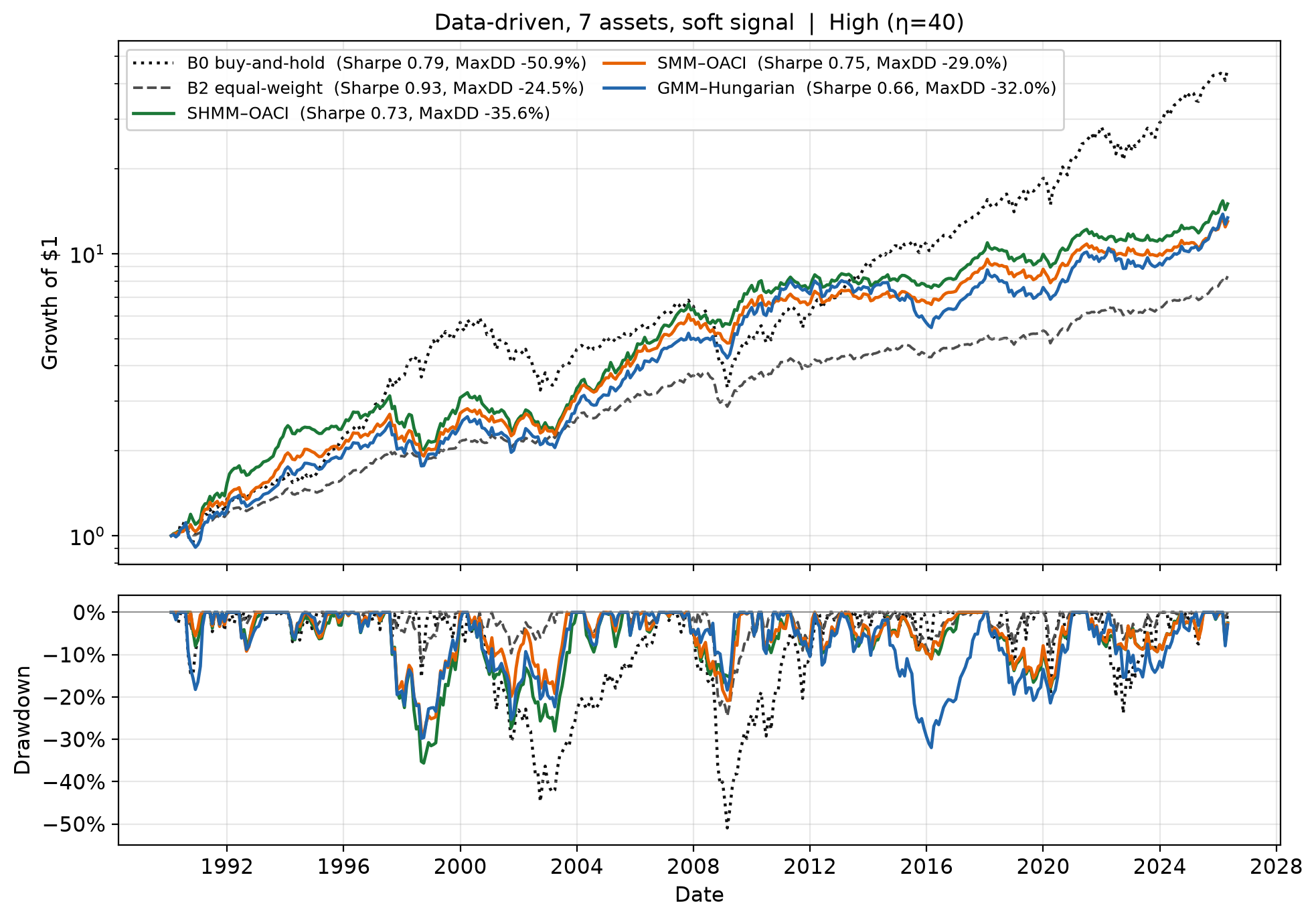}
\caption{Seven-asset portfolios with high growth emphasis ($\eta=40$). Growth of
\$1 on a log scale and drawdown for the strategies and benchmarks in
Figure~\ref{fig:datadriven-eight-path-low}.}
\label{fig:datadriven-eight-path-high}
\end{figure}
\section{Matched comparisons of regime-conditioned allocations}
\label{app:allocation-controls}

At month-end $t$, both updates use the same expanding history of
seven-asset returns $r_u$, $u\le t$, starting with 60 observations.
For regime $k$, the same-month rule weights $r_u$ by
$\tilde g_u(k)$ and the next-month rule by $\tilde g_{u-1}(k)$.
Each regime distribution is shrunk toward the pooled distribution
with weight $24/(24+n_{\mathrm{eff}})$, where
$n_{\mathrm{eff}}=1/\sum_u v_u^2$ for normalized score weights $v_u$.
Both rules maximize the annualized Sharpe ratio ($r_f=0$) within each sleeve and
combine the four regime portfolios using $\tilde g_t$ for month $t+1$.\\
The pooled control uses the same historical seven-asset return vectors
as the two sleeve strategies, giving every month equal weight regardless
of its regime scores. At each rebalance, it estimates one vector of asset
means and one covariance matrix from this history, then uses them to
choose a single maximum-Sharpe portfolio. All three rules use long-only,
fully invested portfolios with a 40\% asset cap and pay five basis
points per dollar traded after drift. Table~\ref{tab:allocation-controls}
reports paired Sharpe differences over 375 invested months. We obtain
pointwise 95\% intervals by jointly resampling monthly net returns
with a stationary bootstrap (2,000 draws, mean block length 12 months).

\begin{table}[H]
\centering
\caption{Differences in annualized Sharpe ratios ($r_f=0$) for the maximum-Sharpe portfolio updates, February 1995-April 2026.}
\label{tab:allocation-controls}
\begin{tabular}{@{}lrr@{}}
\toprule
Comparison & Sharpe difference & 95\% interval \\
\midrule
Next-month minus same-month sleeves & $-0.08$ & $[-0.25,0.05]$ \\
Next-month sleeves minus pooled control & $+0.09$ & $[-0.02,0.22]$ \\
\bottomrule
\end{tabular}
\end{table}

\end{document}